\documentclass[conference]{IEEEtran}
\IEEEoverridecommandlockouts

\usepackage{cite}
\usepackage{amsmath,amssymb,amsfonts}
\usepackage{algorithmic}
\usepackage{graphicx}
\usepackage{textcomp}
\usepackage{xcolor}

\usepackage{algorithm}
\usepackage{amsthm}
\usepackage{array}
\usepackage[english]{babel}
\usepackage{color, colortbl}
\usepackage{bm}
\usepackage{comment}
\usepackage{dsfont}
\usepackage{enumerate}
\usepackage[leqno,fleqn,intlimits]{empheq}
\usepackage{float}
\usepackage{graphicx} 		
\usepackage[margin=1in]{geometry}
\usepackage[bookmarks=false]{hyperref}
\usepackage[utf8]{inputenc}
\usepackage{lipsum}  
\usepackage{lineno}
\usepackage[framemethod=tikz]{mdframed}
\usepackage{microtype}      	
\usepackage{mathtools}
\usepackage{multicol}
\usepackage{multirow}
\usepackage{nicefrac}       	
\usepackage{rotating}
\usepackage{stmaryrd}
\usepackage{subcaption}
\usepackage{tabularx}
\usepackage{textcomp}
\usepackage{tikz}
\usepackage{upgreek}
\usepackage{url}            		
\usepackage{wrapfig}
\usepackage{xcolor}

\usepackage[square,numbers,sectionbib]{natbib}
\usepackage{chapterbib}

\newtheorem{theorem}{Theorem}
\newtheorem{proposition}{Proposition}

\DeclarePairedDelimiter\abs{\lvert}{\rvert}
\newcommand{\norm}[1]{\left\lVert#1\right\rVert}

\DeclareMathAlphabet\mathbfcal{OMS}{cmsy}{b}{n}
\DeclareMathOperator*{\argmax}{arg\,max}
\DeclareMathOperator*{\argmin}{arg\,min}

\definecolor{col1}{HTML}{000000}
\definecolor{col2}{HTML}{000000}
\definecolor{col3}{HTML}{000000}
\definecolor{col4}{HTML}{000000}

\usetikzlibrary{shadows}
\definecolor{shadowcolor}{rgb}{0,.5,.5}
\newmdenv[shadow=true,shadowcolor=shadowcolor,font=\sffamily,rightmargin=8pt]{shadedbox}

\definecolor{Gray}{gray}{0.9}
\definecolor{LightCyan}{rgb}{0.88,1,1}

\def\BibTeX{{\rm B\kern-.05em{\sc i\kern-.025em b}\kern-.08em
    T\kern-.1667em\lower.7ex\hbox{E}\kern-.125emX}}

\begin{document}

\title{Interpretable Visualization and Higher-Order Dimension Reduction for ECoG Data\\
\thanks{$*$ denotes equal contribution. 1-3 denote additional affiliations with 1. Dept. of Statistics, Rice University, 2. Dept. of Computer Science, Rice University, and 3. Jan and Dan Duncan Neurological Research Institute, Baylor College of Medicine.}
}
\author{\IEEEauthorblockN{Kelly Geyer$^*$}
\IEEEauthorblockA{\textit{Dept. of Mathematics \& Statistics} \\
\textit{Boston University}\\
Boston, MA \\
klgeyer@bu.edu}
\and
\IEEEauthorblockN{Frederick Campbell$^*$}
\IEEEauthorblockA{
\textit{Microsoft}\\
Redmond, WA \\
frcampbe@microsoft.com}
\and
\IEEEauthorblockN{Andersen Chang}
\IEEEauthorblockA{\textit{Dept. of Statistics} \\
\textit{Rice University}\\
Houston, TX \\
atc7@rice.edu}
\and
\IEEEauthorblockN{John Magnotti}
\IEEEauthorblockA{\textit{Dept. of Neurosurgery} \\
\textit{Baylor College of Medicine}\\
Houston, TX \\
magnotti@bcm.edu}
\and
\IEEEauthorblockN{Michael Beauchamp}
\IEEEauthorblockA{\textit{Dept. of Neurosurgery} \\
\textit{Baylor College of Medicine}\\
Houston, TX \\
michael.beauchamp@bcm.edu}
\and
\IEEEauthorblockN{Genevera I. Allen$^{1,2,3}$}
\IEEEauthorblockA{\textit{Dept. of Electrical and Computer Engineering} \\
\textit{Rice University}\\
Houston, TX \\
gallen@rice.edu}
}
\maketitle

\begin{abstract}
ElectroCOrticoGraphy (ECoG) technology measures electrical activity in the human brain via electrodes placed directly on the cortical surface during neurosurgery. 
Through its capability to record activity at a fast temporal resolution, ECoG experiments have allowed scientists to better understand how the human brain processes speech. 
By its nature, ECoG data is difficult for neuroscientists to directly interpret for two major reasons. 
Firstly, ECoG data tends to be large in size, as each individual experiment yields data up to several gigabytes. 
Secondly, ECoG data has a complex, higher-order nature.
After signal processing, this type of data may be organized as a 4-way tensor with dimensions representing trials, electrodes, frequency, and time. 
In this paper, we develop an interpretable dimension reduction approach called Regularized Higher Order Principal Components Analysis, as well as an extension to Regularized Higher Order Partial Least Squares, that allows neuroscientists to explore and visualize ECoG data.  
Our approach employs a sparse and functional Candecomp-Parafac (CP) decomposition that incorporates sparsity to select relevant electrodes and frequency bands, as well as smoothness over time and frequency, yielding directly interpretable factors. 
We demonstrate the performance and interpretability of our method with an ECoG case study on audio and visual processing of human speech.
\end{abstract}


\section{Introduction}\label{sec::introduction}
ElectroCOrticoGraphy (ECoG) is a technique for measuring the electrical activity of a brain over time, among several locations in the brain. 
Because it can record brain activity at a higher spatiotemporal frequency and with less noise than previous technologies, ECoG studies have helped scientists understand better than before how the human brain functions for tasks such as speech recognition and processing. 
In the typical experimental setting involving ECoG, a stimulus such as an image or noise is presented to the subject and the responding activity in their brain is measured. 
Within each trial, a subject is exposed to a single stimulus, and the electrical activity of the subject's brain is measured by electrodes placed on the surface of the brain. 
Fourier transforms are then used to obtain spectrograms of brain activity over time for each electrode in each trial.
ECoG data naturally give rise to a complex, high-dimensional, higher-order structure which has dependent measurements; the resulting data structures have thousands of covariates, and their files sizes usually exceeds 3-4 GB. Thus, without proper techniques, analyzing and interpreting ECoG data from the large amount of temporal-spatial information collected can be challenging from both a computational and analytical standpoint.

Several approaches have been used in the neuroscience literature to analyze and interpret ECoG data. 
Researchers may typically analyze a small number of electrodes \citep{kubanek2009decoding}, omitting or condensing down the number of variables and observations substantially. 
Additionally, they may process the data using procedures that condense a large number of frequencies into several frequency bands \citep{pei2011decoding}. 
Omission makes an unwieldy data set manageable, but there is work that suggests condensing the data to frequency bands is an oversimplification \citep{gaona2011nonuniform}. 
Another strategy for decoding ECoG data is the usage of classical multivariate statistical methods such as principal component analysis (PCA) \citep{kellis2010decoding} or forward selection \citep{liang2012decoding}. 
In general, classical multivariate analysis methods are commonly used in neuroimaging studies \citep{krishnan2011partial,delorme2004eeglab}. 
The primary disadvantage of these methods is that they require flattening of the higher-order structure of ECoG data in to matrix form, as the merging of different dimensions causes the results to lose much of their scientific interpretability.
Computational feasibility can be a challenge for these methods as well, since representing higher-order ECoG data as a 2 dimensional matrix necessitates the estimation of an astronomically large number of parameters; this can make several popular dimension reduction and classification techniques, such as linear discriminant analysis (LDA), partial least squares (PLS), and logistic regression, computationally infeasible.

A more sophisticated approach for analyzing ECoG data is the use of tensor decomposition methods. 
Despite the complex nature of ECoG data, its representation can be simplified and ultimately be described by four modalities: $i.)$ trial for each stimulus presented, $ii)$ electrodes, $iii)$ frequency resulting from measurement of electrical activity, and $iv)$ recording time.
The modality measurements of ECoG are consistent, despite potential differences between various experimental designs and settings; this means that in general ECoG data may be represented by a four-dimensional tensor, i.e. in the form $\mathbfcal{X} \in \mathbb{R}^{n \times p \times q \times r}$. 
Tensor decompositions greatly reduce the number of estimated parameters, by returning one or more factor vectors for each mode. 
\citet{cichocki2013tensor} and \citet{cichocki2015tensor} suggest that tensors decomposition can improve the analysis of data with spatial-temporal structure, by finding patterns in each of the modes separately. 
Tensor decomposition methods have been used to analyze ECoG in many previous studies \citep{eliseyev2012l1, zhao2013higher, zhao2013kernel, zhao2013kernelization, eliseyev2013recursive,zhao2014multilinear}. 
While these methods preserve the higher-order nature of ECoG data, the results can still be difficult for scientists to interpret, as one can not easily perform feature selection using the results from a basic tensor decomposition.


In this paper, we introduce Regularized Higher-Order Principal Component Analysis ($\rho$-PCA) as a method of dimension reduction for ECoG data, as well as an extension to Regularized Higher Order Partial Least Squares ($\rho$-PLS). 
Our method aims to improve the interpretability of the results of tensor decompositions on ECoG data. 
To do this, we use regularized tensor factorization techniques, adopting the work done in previous papers for sparse higher-order PCA \citep{allen2012sparse, Allen:2019, allen2013multi} and building on the methodology presented in those papers to be specifically applied to ECoG data. 
In particular, $\rho$-PCA returns sparse tensor estimates along the electrode and frequency modes, which in turn helps to automatically identify the important features along those modalities by detecting the non-zero elements of the tensors. 
The framework of $\rho$-PCA also allows for other structures to be imposed on the estimated tensors, such as smoothness over time. 
We apply this new methodology to ECoG data from a speech perception experiment, in which we use $\rho$-PCA to make predictions about the stimuli the patient is experiencing though the observed brain response. 
We show that our method is competitive with other approaches currently used in the literature for analyzing ECoG data, both in terms of classification accuracy and in terms of computational performance. 
We also demonstrate how one can easily interpret and visualize the results of $\rho$-PCA through analyzing a specific trial from the aforementioned ECoG data set.


\section{Tensor Decomposition for ECoG Analysis} \label{sec:methodology}
In this section, we describe $\rho$-PCA mathematically. 
We refer to the dimensions of a tensor as its modes, and adopt the notation from \citet{Kolda1}.
We denote tensors as $\mathbfcal{X}$, matrices as $\bm{X}$, vectors as $\bm{x}$, and scalars as $x$. 
A tensor may be multiplied by a matrix or vector by the $d$-th mode, using the operation $\times_d$.
We can flatten a tensor into a matrix along the $d$-th mode, denoted by $\bm{X}_{(d)}$.
We define vector norms $\norm{\bm{x}}_1 = \sum_{i=1}^{N} \abs{x_i}$, $\norm{\bm{x}}_2^2 = \sum_{i=1}^N \abs{x_i}^2$, and $\norm{\bm{x}}_{\bm{\Omega}}^2 = \bm{x}^T \bm{\Omega} \bm{x}$, where $\bm{\Omega} \in \mathbb{R}^{N \times N}$.
The Frobenius norm for a matrix is $\norm{\bm{X}}_F^2 = \sum_{i=1}^{N} \sum_{j=1}^{M} \abs{x_{ij}}^2$; this may be generalized for a tensor.
The trace operation is $\text{Tr}\left(\bm{X}\right) = \sum_{i=1}^{n}x_{ii}$ for $X \in \mathbb{R}^{n \times n}$.
The soft-thresholding function is defined as $S(x,\lambda) = sign(x) \cdot \left(\abs{x} - \lambda \right)_+$. 

\subsection{Building the Methodology of $\rho$-PCA}
Here, we describe how we derive $\rho$-PCA from the tensor decomposition called sparse higher-order principal components analysis \citep{allen2012sparse}, as well the motivation behind our choice of regularization for each modality.
As previously stated, an ECoG recording can be represented as a 4-way tensor, $\mathbfcal{X} \in \mathbb{R}^{n \times p \times q \times r}$.
In this tensor, there are $n$ trials, $p$ electrodes, $q$ frequency bands, and $r$ time stamps.
$\rho$-PCA decomposes the ECoG tensor into $K$ sets of sparse, nearly independent vectors for each modality, that sequentially correspond to the strongest patterns in each mode.

Sparse higher-order principal components analysis is based upon CP-decomposition \citep{allen2012sparse}.
The CP-decomposition of this tensor results in the sum of $K$ outer products of the component factors:
\vspace{1pt}
\begin{equation} \label{eqn::cp-decomp}
\bm{\mathcal{X}} \approx \sum_{k=1}^{K} d_k \cdot \bm{u}_k \circ \bm{v}_k \circ \bm{w}_k \circ \bm{t}_k
\end{equation}
\vspace{1pt}
\noindent where $\{\bm{u}_k\}_{k=1}^K \in \mathbb{R}^n$, $\{\bm{v}_k\}_{k=1}^K \in \mathbb{R}^p$, $\{\bm{w}_k\}_{k=1}^K \in \mathbb{R}^q$, and $\{\bm{t}_k\}_{k=1}^K \in \mathbb{R}^r$ are factor vectors for trials, electrodes, frequencies and time, respectively.
All of these vectors are standardized to have a $\ell_2$-norm of one, and the weights are absorbed into the constants $\{d_k\}_{k=1}^K \in \mathbb{R}$. 
These factor vectors can be estimated using the optimization problem for CP-decomposition, which seeks a low-rank solution. 
The optimization problem below, for example, is used to find a rank $K = 1$ solution \footnote{We use $\rho$-PCA to find $K \geq 1$ sets of factor vectors for each modality.}:
\vspace{1pt}
\begin{equation} \label{eqn::cp-decomp-v2}
\begin{aligned}
& \underset{\bm{u, v, w, t}}{\text{minimize}}
& & \norm{\bm{\mathcal{X}} - d \cdot \bm{u} \circ \bm{v} \circ \bm{w} \circ \bm{t}}_F^2  \\
& \text{subject to}
& & \bm{u}^T\bm{u} \leq 1, \bm{v}^T\bm{v} \leq 1, \\ &&& \bm{w}^T\bm{w} \leq 1, \bm{t}^T\bm{t} \leq 1.   
\end{aligned}
\end{equation}
\vspace{1pt}
The solution of the CP-decomposition may be found using the Alternating Least Squares (ALS) algorithm.  
However, this can take many iterations to converge, and it is not guaranteed to converge to a global minimum or stationary point \citep{Kolda1}. 
Instead, we use the Tensor Power Algorithm (TPA) for $\rho$-PCA, proposed by \citet{allen2012sparse}. 
There are several benefits to using a TPA-based approach over ALS.
Firstly, TPA always converges to a point that is stationary, while ALS may not converge to one.
Secondly, TPA is a greedy approach in that the first factors computed will explain the most variance in the data, thus yielding component vectors that explain sequentially decreasing amounts of variance within the data.
TPA enables one to computer fewer components overall, while potentially capturing the most variance. 
Lastly, the deflation step of TPA enforces pseudo-independence among each set of component vectors. 
To proceed with the TPA approach, we recast Eq. \eqref{eqn::cp-decomp-v2} using results from \citet{kolda2006multilinear}, and obtain Eq. \eqref{eqn:opteq2}:
\begin{equation} \label{eqn:opteq2}
\begin{aligned}
&\underset{\bm{u},\bm{v},\bm{w},\bm{t}}{\text{maximize }} && \mathbfcal{X} \times_1  \bm{u} \times_2 \bm{v} \times_3 \bm{w} \times_4 \bm{t} \\
 &\text{subject to } &&  \bm{u}^T \bm{u} \le 1, \bm{v}^T \bm{v}  \le1, \\ &&& \bm{w}^T \bm{w} \le1, \bm{t}^T\bm{t}  \le 1.
\end{aligned}
\end{equation}
\vspace{1pt}
\noindent 
It is simple to verify that Eqns. \eqref{eqn::cp-decomp-v2} and \eqref{eqn:opteq2} are equivalent \citep{allen2012sparse}. 
Notice that Eq. \eqref{eqn:opteq2} is separable in the factors, and enables iterative block-wise optimization.
That is, the objective function is maximized with respect to one of the variables $\bm{u}$,  $\bm{v}$, $\bm{w}$ or $\bm{t}$, while holding all others fixed for each iteration.

To enhance the interpretability of results, we add regularization penalties and constraints to the optimization problem in Eq. \eqref{eqn:opteq2}. 
Our methodology is an extension of regularized tensor factorization by \citet{allen2011sparse}, with penalties specific to each mode of the ECoG data. 
This tensor factorization decomposes an $N$-way tensor into $N$ sets of component factors. 
General convex penalties and constraints can then be applied to each of these component vectors separately \citep{allen2011sparse}. 
We propose to use the following types of regularization for each ECoG mode based on their known structures: 
\begin{enumerate}[i)]\itemsep0pt 
    \item \textbf{Electrodes:}  We propose to use an $\ell_1$-norm penalty, denoted as $\norm{x}_1$, to select the most relevant electrodes for specific tasks.
    \item \textbf{Frequency:} We propose to select relevant frequencies, and smooth them to yield more interpretable results.  
    To this end, we employ the sparse and functional penalties from \citet{Allen:2019}.
    The frequency mode is penalized by both an $\ell_1$-norm penalty to select the most relevant frequencies, and a difference matrix $\bf{\Omega}_w$ to smooths over frequency values.
    Smoothing over frequencies removes noise observed between similar frequencies.
    \item \textbf{Time:} We propose a smoothing penalty over time to yield more interpretable time factors by removing noise from observations close in time.
    We penalize the time factors with a difference matrix $\bm{\Omega}_t$.
\end{enumerate}

\begin{figure}[t!] 
   \centering
    \includegraphics[width=\linewidth]{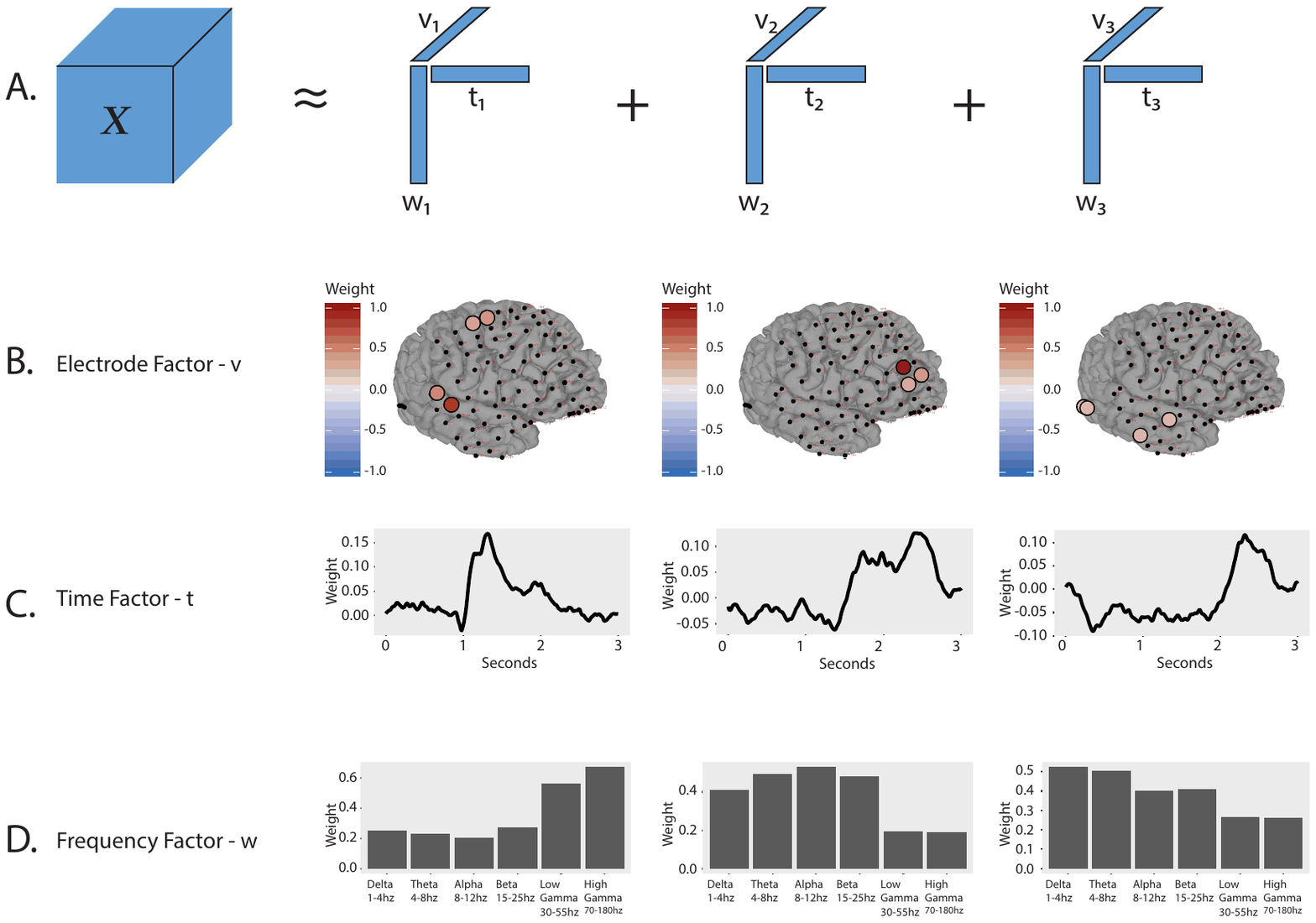}
    \caption{Clear \& noisy audio classification for patient YAK using $\rho$-PLS. 
    A.) ECoG tensor $\mathcal{X}$ is decomposed into three components. 
    Each component has an electrode factor, a time factor and a frequency factor. 
    B.) The electrode factors highlight which electrodes exhibit a differential response to the two stimuli classes. 
    C.) The time factors show when there is a differential response to the two stimuli classes. 
    D.) The frequency factors communicate which frequency bands have a differential response to the stimuli.}
    \label{fig:decomposition_figure}
\end{figure}

Eq. \eqref{eqn:opteq2} is the result of applying these penalities to the rank-one tensor model, in order to yield the optimization problem that defines $\rho$-PCA:
\begin{equation} \label{eqn::p-pca-v2}
\begin{aligned}
 \underset{\bm{u, v, w, t}}{\text{maximize}} & & \bm{\mathcal{X}} \times_1 \bm{u} \times_2 \bm{v} \times_3 \bm{w} \times_4 \bm{t} \\ && - \lambda_v\norm{\bm{v}}_1 - \lambda_w \norm{\bm{w}}_1 \\
 \text{subject to} & & \bm{u}^T\bm{u} \leq 1, \bm{v}^T\bm{v} \leq 1, \\ && \bm{w}^T(\bm{I} + \alpha_w\bm{\Omega_w})\bm{w} \leq 1, \\ && \bm{t}^T(\bm{I}  +  \alpha_t\bm{\Omega_t})\bm{t} \leq 1.
\end{aligned}
\end{equation}
\vspace{1pt}
\noindent The matrices $\bm{\Omega}_w$ and $\bm{\Omega}_t$ are 2nd-order difference matrices, and ultimately smooth over notable changes in signal over time and frequencies values.
The constants $\lambda_v, \lambda_w \geq 0$ control the amount of sparsity applied to the electrode and frequency factors.
Increasing these values will greatly shrink the factors, as well as set some to 0, making them less significant.
The constants $\alpha_w, \alpha_t \geq 0$ control the amount of smoothing over frequency and time. 
Increasing these values makes the elements of the factor vectors smoother and more homogeneous.
Additionally, we define $\bm{S}_w = \bm{I}_q + \alpha_w\bm{\Omega}_w \succ 0$, $L_w$ as the largest eigenvalue of $\bm{S}_w$, and $\bm{S}_t = \bm{I}_r + \alpha_t\bm{\Omega}_t \succ 0$. 
Notably, the $\ell_1$-norm penalties in Eq. \eqref{eqn::p-pca-v2} enable us to perform automatic data-driven feature selection along the electrode and frequency modalities while estimating the factors in $\rho$-PCA. 
They interpretable results, as one can easily identify the most important electrodes and frequencies in an ECoG data set simply by finding the non-zero elements of the factors.

Figure \ref{fig:decomposition_figure} contains $\rho$-PCA factors for patient YAK, and shows the estimated tensor decomposition along three modalities. 
The sparsity penalty of the electrode mode allow us to identify 3-4 influential electrodes for each tensor component, as well as evaluate the corresponding time and frequency factors associated with those particular electrodes. 

Alg. \ref{alg::p-pca} is a computationally efficient algorithm for computing $\rho$-PCA, and it is an extension of TPA \citep{allen2012sparse}.
It utilizes a coordinate-wise update schema in which each mode is updated individually, while holding the other ones constant, in an iterative manner until convergence is reached. 
As with TPA in general, Alg. \ref{alg::p-pca} is guaranteed to converge to a local maximum:
\begin{theorem}
\label{thm::bloc-sol-p-pca}
Each factor-wise block coordinate-wise update of $\rho$-PCA monotonically increases the objective that when iterated, converge to a local maximum of the optimization problem \eqref{eqn::p-pca-v2}.
\end{theorem}

\addtocounter{theorem}{-1}

In addition to Theorem \ref{thm::bloc-sol-p-pca}, there are two more significant advantages of using TPA for $\rho$-PCA.
Firstly, the $\ell_2$-norm and $\norm{\cdot}_{\bm{\Omega}}$ are equal to one or zero to avoid degenerate solutions \citep{allen2012regularized}.
Secondly, the block coordinate solutions are simple and closed-form, making each update is inexpensive to compute \citep{allen2012regularized}.

Hyperparameter selection for $\rho$-PCA can be performed using a nested Bayesian Information Criterion schema \citep{lee2010biclustering, allen2011sparse}.
A full description of hyperparameter selection for $\rho$-PCA is presented in Alg. 3 in Appendix B.

\begin{algorithm}[t!]
\caption{$\rho$-PCA Algorithm via the Tensor Power Method}\label{alg::p-pca}
\begin{small}
\begin{enumerate}\itemsep0pt 
    \item Initialization:
    \begin{enumerate}
        \item Set $\hat{\mathbfcal{X}} = \mathbfcal{X}$.
        \item Pre-compute $\bm{S}_w$ and $\bm{S}_t$.
        \item Initialize $\left\{\left(\hat{d}_k, \hat{\bm{u}}_k, \hat{\bm{v}}_k, \hat{\bm{w}}_k, \hat{\bm{t}}_k\right)\right\}_{k=1}^{K}$. 
    \end{enumerate}
    \item For $k = 1, \ldots, K$:
    \begin{enumerate}[(a)]\itemsep0pt
    
        \item Repeat until convergence:
        \begin{enumerate}[i.]\itemsep0pt
        
            \item Update trial component estimate $\hat{\bm{u}}_k$,
            \vspace{0pt}
            \begin{equation} \label{eqn::u-update}
            \begin{split}
                \tilde{\bm{u}}_k & = \hat{\mathbfcal{X}} \times_2 \hat{\bm{v}}_{k-1} \times_3 \hat{\bm{w}}_{k-1} \times_4 \hat{\bm{t}}_{k-1}, \\ \quad
                \hat{\bm{u}}_k & = 
              	\begin{cases}
            		\frac{\tilde{\bm{u}}_k}{\norm{\tilde{\bm{u}}_k}_2}, & \norm{\tilde{\bm{u}}_k}_2 > 0, \\
            		0, & otherwise
            	\end{cases}
            \end{split}
            \end{equation}
            \vspace{0pt}
            \item Update electrode component  estimate $\hat{\bm{v}}_k$,
            \vspace{0pt}
            \begin{equation} \label{eqn::v-update}
            \begin{split}
                \tilde{\bm{v}}_k & = S(\hat{\mathbfcal{X}} \times_1 \hat{\bm{u}}_{k-1} \times_3 \hat{\bm{w}}_{k-1} \times_4 \hat{\bm{t}}_{k-1}, \lambda_v), \\ \quad
                \hat{\bm{v}}_k & = 
              	\begin{cases}
            		\frac{\tilde{\bm{v}}_k}{\norm{\tilde{\bm{v}}_k}_2}, & \norm{\tilde{\bm{v}}_k}_2 > 0, \\
            		0, & otherwise
            	\end{cases}
            \end{split}
            \end{equation}
            \vspace{-10pt}
            \item Update frequency component estimate $\hat{\bm{w}}_k$,
            \vspace{0pt}
            \begin{equation} \label{eqn::w-update}
                \begin{split}
                \tilde{\bm{w}}_k &= \text{prox}_{\frac{\lambda_w}{L_w} \norm{\cdot}_1} \left(\hat{\bm{w}}_{k-1} + \right. \\ 
                & \left. L_w^{-1} \left( \hat{\mathbfcal{X}} \times_1 \hat{\bm{u}}_{k-1} \times_2 \hat{\bm{v}}_{k-1} \times_4 \hat{\bm{t}}_{k-1}   - \right. \right. \\  & \left. \left. \hspace{20pt} \bm{S}_w \hat{\bm{w}}_{k-1}\right) \right) \\
                \hat{\bm{w}}_k & = 
              	\begin{cases}
            		\frac{\tilde{\bm{w}}_k}{\norm{\tilde{\bm{w}}_k}}_{\bm{S}_w}, & \norm{\tilde{\bm{w}}_k}_{\bm{S}_w} > 0, \\
            		0, & otherwise		
            	\end{cases}
            	\end{split}
            \end{equation}
            \vspace{-10pt}
            \item Update time component estimate $\hat{\bm{t}}_k$,
            \vspace{0pt}
            \begin{equation}  \label{eqn::t-update}
            \begin{split}
            \tilde{\bm{t}}_k & = \bm{S}_t^{-1} \left( \hat{\mathbfcal{X}} \times_1 \hat{\bm{u}}_{k-1} \times_2 \hat{\bm{v}}_{k-1} \times_3 \hat{\bm{w}}_{k-1} \right), \\ \quad
            \hat{\bm{t}}_k & = 
            	\begin{cases}
            		\frac{\tilde{\bm{t}}_k}{\norm{ \tilde{\bm{t}}_k }_{\bm{S}_t}}, & \norm{ \tilde{\bm{t}}_k }_{\bm{S}_t} > 0, \\
            		0, & otherwise
            	\end{cases}
            	\end{split}
            \end{equation}
            \vspace{-10pt}
        \end{enumerate}
        \item Deflate to encourage pseudo-independence between sets of $K$ component vectors,
        \vspace{0pt}
            \begin{equation} \label{eqn::deflation}
            \begin{split}
                \hat{d}_k & = \hat{\mathbfcal{X}} \times_1 \hat{\bm{u}}_k \times_2 \hat{\bm{v}}_k \times_3 \hat{\bm{w}}_k \times_4 \hat{\bm{t}}_k, \\ \quad
                \hat{\mathbfcal{X}}&  = \hat{\mathbfcal{X}} - \hat{d}_k \cdot \hat{\bm{u}}_k \circ \hat{\bm{v}}_k \circ \hat{\bm{w}}_k \circ \hat{\bm{t}}_k
            \end{split}
            \end{equation}
  \end{enumerate}
  \item Return factors $\left\{\left(\hat{d}_k, \hat{\bm{u}}_k, \hat{\bm{v}}_k, \hat{\bm{w}}_k, \hat{\bm{t}}_k\right)\right\}_{k=1}^{K}$.
\end{enumerate}
\end{small}
\end{algorithm}

\subsection{Supervised Setting: $\rho$-PLS}
$\rho$-PCA is easily be extended for supervised learning via Partial Least Squares (PLS) regression.
Traditional PLS finds the direction of the covariates that explains the maximum variance in the response \citep{geladi1986partial}.
Recall the optimization problem for traditional PLS:
\vspace{0pt}
\begin{equation} \label{eqn::pls}
\begin{aligned}
& \underset{\bm{v, w}}{\text{maximize}}
& \text{Cov}\left(\bm{v}^T \bm{X}, \bm{w}^T \bm{Y} \right) := \bm{v}^T \underline{\bm{X}^T \bm{Y}} \bm{w} \\
& \text{subject to}
 & \norm{\bm{v}}_2 = \norm{\bm{w}}_2 = 1, \\ 
 & &  \bm{v}^T\bm{v} = \bm{I}_p, \bm{w}^T\bm{w} = \bm{I}_q
\end{aligned}
\end{equation}
\vspace{0pt}
\noindent 
where $\bm{X} \in \mathbb{R}^{n \times p}$, $\bm{Y} \in \mathbb{R}^{n \times q}$, $\bm{v} \in \mathbb{R}^p$, and $\bm{w} \in \mathbb{R}^q$. 
The coefficient vectors have a 2-norm of one, and are constrained to be orthogonal. 
We can easily extend Eq. \eqref{eqn::pls} for higher-order arrays, as well as include regularization, while dropping the orthogonality constraints typical in CP decompositions.
In turn, the traditional covariate tensor $\mathbfcal{X}$ is replaced by the covariance tensor $\mathbfcal{X} \times_1 \bm{y}$, where $\bm{y} \in \{0,1\}^{n}$, is the response.  
The optimization problem in Eq. \eqref{eqn::p-pls} defines the $\rho$-PLS solution:
\vspace{0pt}
\begin{equation} \label{eqn::p-pls}
\begin{aligned}
& \underset{\bm{v, w, t}}{\text{maximize}}
&   \left(\underline{\mathbfcal{X} \times_1 \bm{y}}\right) \times_1 \bm{v} \times_2 \bm{w} \times_3 \bm{t} \\ & & - \lambda_v\norm{\bm{v}}_1 - \lambda_w \norm{\bm{w}}_1 \\
& \text{subject to}
& \bm{w}^T(\bm{I}_q + \alpha_w\bm{\Omega_w})\bm{w} \leq 1, \\
&& \bm{v}^T\bm{v} \leq 1, \text{ \& } \bm{t}^T(\bm{I}_r  +  \alpha_t\bm{\Omega_t})\bm{t} \leq 1  
\end{aligned}
\end{equation}
\vspace{0pt}
Alg. \ref{alg::p-pls} illustrates the implementation of $\rho$-PLS.
Overall, $\rho$-PLS reduces the ECoG tensor data into a reduced form, based on the class separation. 
Following this, the reduced data may be plugged into any supervised learning model. 
We use a nested $K$-fold cross-validation scheme to find optimal hyperparameters for $\rho$-PLS, explained in Alg. 4 in Appendix B.

\begin{algorithm}[t!]
\caption{$\rho$-PLS Algorithm: Extension of $\rho$-PCA for supervised dimension reduction and prediction.}\label{alg::p-pls}
\begin{small}
\begin{enumerate}\itemsep0pt 
    \item Initialization: 
    \begin{enumerate}[(a)]\itemsep0pt 
        \item Define the covariates $\mathbfcal{X} \in \mathbb{R}^{n \times p \times q \times r}$, and response 
        $\bm{y} \in \{0,1\}^{n}$
        \item Center each column of $\bm{y}$, denoted as $\bar{\bm{y}}$.
    \end{enumerate}
    \item Compute the covariance tensor,
    \vspace{0pt}
    \begin{equation*}
        \mathbfcal{Z} = \mathbfcal{X} \times_1 \bar{\bm{y}} \in \mathbb{R}^{p \times q \times r}
    \end{equation*}
    \vspace{-10pt}
    \item Estimate the component vectors,
    \vspace{0pt}
    \begin{equation*}
    \begin{split}
        &\left[\{\hat{\bm{m}}_k\}_{k=1}^{K}, \{\hat{\bm{v}}_k\}_{k=1}^{K}, \{\hat{\bm{w}}_k\}_{k=1}^{K}, \{\hat{\bm{t}}_k\}_{k=1}^{K}\right] =  \\ 
       & \rho \text{-PCA}\left(\mathbfcal{Z}, K, \bm{\lambda}\right) 
    \end{split}
    \end{equation*}
    where $\bm{\lambda} = \{\left(\lambda_v, \lambda_w, \alpha_w, \alpha_t\right)_k\}_{k=1}^K$ are the regularization parameters.
    \item Reduce the dimensionality of the covariates for each $k \in \{1, \ldots, K\}$,
    \vspace{0pt}
    \begin{equation*}
        \tilde{\bm{x}}_k = \mathbfcal{X} \times_2 \hat{\bm{v}}_k \times_3 \hat{\bm{w}}_k \times_4 \hat{\bm{t}}_k  \in \mathbb{R}^{n}
    \end{equation*}
    Concatenate the $\tilde{\bm{x}}_k$'s to form $\tilde{\bm{X}} = \left[\tilde{\bm{x}}_1, \ldots, \tilde{\bm{x}}_K\right] \in \mathbb{R}^{n \times K}$
    \vspace{3pt}
    \item Return the reduced data $\Tilde{\bm{X}}$ and component vectors $\left[\{\hat{\bm{m}}_k\}_{k=1}^{K}, \{\hat{\bm{v}}_k\}_{k=1}^{K}, \{\hat{\bm{w}}_k\}_{k=1}^{K}, \{\hat{\bm{t}}_k\}_{k=1}^{K}\right]$.
    \begin{enumerate}[(a)]\itemsep0pt 
        \item Build model to predict response $\bm{Y}$ from $\Tilde{\bm{X}}$.
        \item Visualize and interpret data, as described in Table \ref{tab::interpretation_table}.
    \end{enumerate}
\end{enumerate}
\end{small}
\end{algorithm}

\begin{table}[t!]
\centering
\caption{Summary of visualizations that can potentially be created from the modes of $\rho$-PLS.}
\begin{small}
\label{tab::interpretation_table}
\begin{tabularx}{\linewidth}{p{15mm}|p{19mm}|l|p{20mm}}
Factor & Formula  & Dimension     & Visualization     \\
\hline
Node & $\hat{\bm{V}} := \{\hat{\bm{v}}_k\}_{k=1}^{K}$ & ($p \times K$) & Sparsity plots display relevant nodes. \\
\hline
Frequency & $\hat{\bm{W}} := \{\hat{\bm{w}}_k\}_{k=1}^{K}$ & ($w \times K$) & Bar charts summarize the frequencies. \\
\hline
Time & $\hat{\bm{T}} := \{\hat{\bm{t}}_k\}_{k=1}^{K}$ & ($r \times K$) & A time series plot shows the shape of the average activity over time. \\
\hline
Observation & $\mathbfcal{X} \times_2 \hat{\bm{v}} \times_3 \hat{\bm{w}}  \times_4 \hat{\bm{t}}$ & ($n \times 1$ )       &  Scatterplots communicate how well the trials separate.      \\
\hline
 Nodes by Time &$\mathbfcal{Z} \times_2 \hat{\bm{w}}$ & ($p \times t$) & Heatmaps show node activity over time. \\
\hline
 Frequency by Time &$\mathbfcal{Z} \times_1  \hat{\bm{v}} $ & ($q \times t$) &  A time series plot shows the frequency intensity over time. \\
\hline
Node by Freq. &$\mathbfcal{Z} \times_3 \hat{\bm{t}}$ & ($p \times q$) &  Bar charts show important node-frequency combinations.
\end{tabularx}
\end{small}
\end{table}

\section{Case Study: Neural-decoding with ECoG}
\label{sec::results}
We compare $\rho$-PCA and $\rho$-PLS to other popular techniques, on a real electrocorticography (ECoG) data set; we use data first analyzed by \citet{ozker2017double}. 
Our analysis differs significantly, but the data collection and preprocessing procedures are the same; we describe these steps below. 

\subsection{Data and Experimental Design}
We seek to make accurate predictions about the stimuli given a patient's ECoG measurements.
Patients have subdural electrodes surgically implanted, and agree to participate in an experiment in which each they are shown a movie of a woman saying either the word ``Rock" or ``Rain". 
The conditions of the audio-visual speech stimuli are varied by either adding noise to the audio or visual, in all possible combinations.
This results in a total of eight classes of stimuli movies.
For each trial, a patient watches a single video and tries to identify the word stated.

\begin{table*}[t!]
\caption{This table compares the average accuracy, average audio run time, and standard deviations for ten trials, of various classification methods for patient YAH.} 
\label{tab:yah-accuracy}
\centering
\begin{small}
\begin{tabular}{l|l|l|l|l|l}
  \hline
& Model & Word & Visual & Audio & Run Time (sec) \\ 
  \hline
  \multirow{5}{5.5em}{Unsupervised Methods}
  &matrix PCA(X) + LDA & 0.55 (0.09) & 0.91 (0.08) & 0.67 (0.08) & 159.63 (31.27) \\ 
  &matrix ICA(X) + LDA & 0.55 (0.10) & 0.73 (0.15) & 0.60 (0.11) & 173.76 (10.70)\\ 
  &CP-ALS(X) + LDA & 0.60 (0.07) & 0.85 (0.07) & 0.65 (0.09) & 143.56 (36.04) \\ 
  &Tucker-ALS(X) + LDA & 0.47 (0.10) & 0.62 (0.15) & 0.50 (0.16) & 78.82 (17.50)\\ 
  &$\rho$-PCA(X) + LDA & 0.59 (0.12) & 0.88 (0.09) & 0.66 (0.11) & 1168.54 (424.44) \\ 
  \hline
  \multirow{7}{5.5em}{Supervised Methods} 
  &matrix SVM(X,y) & 0.71 (0.10) & 0.94 (0.05) & 0.90 (0.07) & 175.31 (42.75) \\  
  &matrix PLS (X) + LDA &  0.65 (0.12) & 0.91 (0.06) & 0.66 (0.11) & 14.90 (1.59)\\ 
  &CP-ALS-PLS (X) + LDA & 0.67 (0.08) & 0.91 (0.05)	& 0.85 (0.10) & 45.61 (3.21) \\ 
  &Tucker-ALS-PLS(X) + LDA & 0.50 (0.10) & 0.89 (0.06) & 0.76 (0.06) & 46.21 (5.10) \\ 
  &N-PLS + LDA & 0.57 (0.09) & 0.93 (0.05) & 0.84 (0.04) & 2254.42 (292.17)\\ 
  & HOPLS + LDA & 0.48 (0.07) & 0.91 (0.06) & 0.62 (0.10) & 709.08 (51.91)\\ 
  &$\rho$-PLS + LDA & 0.65 (0.06) & 0.89 (0.05)	& 0.89 (0.08) & 48.21 (67.86) \\ 
\end{tabular}
\end{small}
\end{table*} 

\subsubsection{Preprocessing Data}
The data was processed using the FieldTrip Toolbox in MATLAB 8.5.0 \citep{oostenveld2011fieldtrip,guide1998mathworks} and RAVE \citep{magnotti2020rave}. 
A common average reference \citep{ludwig2009using} was used to remove artifacts from the electrodes. 
The data was epoched into trials lasting three seconds. 
Line noise was removed at 60, 120 and 180 Hz. 
We transformed the data to time-frequency space using the multi-taper method with three Slepian tapers, a frequency window from 10 to 200 Hz, frequency steps of 2 Hz, time steps of 10 ms, temporal smoothing of 200 ms, and frequency smoothing of 10 Hz.
After processing, the resulting data set measures signal power as a function of frequency and time at each electrode.

\subsubsection{Formatting ECoG Tensors}
We concatenate trials, and form an ECoG tensor for each patient.
Each patient’s brain is different and, as a result, it is ill-advised to perform classification across patients.
A typical patient participates in around 150 trials and has about 100 electrodes, 100 frequencies and a duration of 3 seconds (measured as 301 discrete points) yielding a 4D tensor. 
These data sets are between 3-5 GB for each patient. 

\subsection{Pipeline for Analysis}
Our analysis consists of several steps that process, analyze and interpret the data:
\begin{enumerate}\itemsep0pt
\item Center and scale the data by the baseline mean and standard deviation, respectively; these are calculated using the measurements from the first half second of the trial, before any stimuli are presented.
\item Select regularization parameters in a component-wise fashion. 
This is done in a data-driven manner, following either of the procedures Alg. 3 or Alg. 4 in Appendix B.
\item  Run either $\rho$-PCA or $\rho$-PLS for supervised or unsupervised learning.
\item  Reduce the dimensionality of the data for decoding, and use the resulting components to visualize the results. (See Table \ref{tab::interpretation_table}.) 
\item Components from the $\rho$-PCA and $\rho$-PLS methods also be plugged into any general statistical learning model as covariates.
\end{enumerate}

\begin{table}[t!]
\caption{This table compares the average accuracy and standard deviation of $\rho$-PLS + LDA, across each patient.}
\label{tab:sub-acc}
\centering
\begin{tabular}{llll}
  \hline
Subject & Word & Visual & Audio \\ 
  \hline
  YAK & 0.61 (0.08)	& 0.96 (0.06) & 0.58 (0.14)	\\
  YAH & 0.65 (0.06)	& 0.89 (0.05) & 0.89 (0.08) \\
  YBA & 0.59 (0.17)	& 0.99 (0.02) & 0.49 (0.08)	\\
\hline
\end{tabular}
\end{table} 

\subsection{Comparative Study of Multi-Sensory Speech Decoding}
We perform a comparative empirical study to demonstrate that $\rho$-PCA and $\rho$-PLS produce factors that can be used to accurately classify aspects of the stimuli. 
Our goal is to predict the stimuli conditions for trials, given a patient's observed ECoG data. 
We pose this problem as a set of binary classification problems for clean \& noisy visuals, clean \& noisy audio, and the words ``rock" \& ``rain". While $\rho$-PCA and $\rho$-PLS can be used for a multi-class classification problem, we choose to simplify the problems into binary cases because this is more interesting in the context of neuroscience. 

We compare $\rho$-PCA and $\rho$-PLS to several different methods that are common in the neuroscience field for analysis. 
This includes several unsupervised methods such as principal component analysis (matrix PCA) \citep{shimada2017impact}, independent component analysis (matrix ICA) \citep{chao2013mining}, Alternating Least Squares for the Candecomp-Parafac decomposition \citep{TTB_Software} (CP-ALS), and Alternating Least Squares for the Tucker decomposition (Tucker-ALS) \citep{zhao2013kernel,zhao2013kernelization}. 
We also compare our method to supervised methods, including support vector machines (matrix SVM), ordinary partial least squares (matrix PLS) \citep{krishnan2011partial}, CP-ALS with partial least squares (CP-ALS-PLS), Tucker-ALS with partial least squares (Tucker-ALS-PLS), multiway partial least squares \citep{bro1996multiway} (N-PLS), and Higher Order Partial Least Squares \citep{zhao2013higher} (HOPLS); for all of the supervised methods, we treat the stimuli conditions as known.  
To fairly compare each dimension reduction approach, we combine each method with LDA, understanding that other approaches (e.g. SVM) might perform better. 
Also, to be fair to all methods for computational timing purposes, we first select oracle hyperparameters for the $\rho$-PCA, $\rho$-PLS, and SVM approaches.  
To do this, we assign 80\% of trials to a training set, while the remaining 20\% of trials are held out to be used as a test set. 
The selected hyperparameters are then fixed for the rest of the simulation study. 

To compare the different methods of classification, we perform 10 replications of model fitting.
For each replication, we randomly assign 90\% of the trials to the training set, and 10\% to the testing set. 
We train each of the different models using the training set, then estimate the classification accuracy of the methods using the test set. 
The results are our study are shown in Table \ref{tab:yah-accuracy}, which contains the average accuracy for each method, average run time (seconds) for the audio classification problem, and the standard deviations in parentheses.
The accuracy for Word classification is noticeably poor.
Classifying words is challenging with ECoG data since specific parts of the brain process different phonemes.  
It is possible there may not be electrodes on the exact spot in the brain that processes rock vs. rain, and hence the difficulty. 

First, we compare the unsupervised methods in Table \ref{tab:yah-accuracy}.
In general, prediction accuracy for Word and Audio are poor, yet more promising for Visual.
In terms for Visual, the methods with the strongest accuracy are matrix PCA, CP-ALS, and $\rho$-PCA.
The first two of these methods perform similarly, while $\rho$-PCA is notably slower. 
We hypothesize that this may be in part due to the fact that the algorithms for both matrix PCA and CP-ALS have been optimized in MATLAB with regards to the singular value decomposition and tensor operations, while our implementation of $\rho$-PCA has not been.  
Further, because $\rho$-PCA induces sparsity and smoothness, it requires more factors to achieve the same level of classification accuracy (a result well-known in the sparse PCA literature \citep{allen2011sparse}), thus further leading to longer run times.  
Lastly, while $\rho$-PCA has a slower runtime, we gain interpretability as a trade-off; CP-ALS and $\rho$-PCA can provide the interpretations from Table \ref{tab::interpretation_table}, and only $\rho$-PCA yields directly interpretable factors.
It is worthy to note that $\rho$-PCA is much slower than $\rho$-PLS in Table \ref{tab:yah-accuracy}.
This is largely due to the dimension reduction while calculating the covariace $\mathbfcal{X} \times_1 \bm{y}$, before running the $\rho$-PLS method.

Next, we compare the supervised methods in Table \ref{tab:yah-accuracy}, which produce promising prediction accuracy across Visual and Audio.
Several methods perform strongly in one or two of the categories, yet do poorly in the other.
Methods that appear to have reasonable prediction accuracy across the Visual and Audio tasks are the matrix SVM, CP-ALS-PLS, and $\rho$-PLS.
Matrix SVM has the highest prediction accuracy across all categories.
However, unlike the other front-runners, it doesn't offer interpretable solutions as it flattens the tensor. 
Also, Matrix SVM is much slower than CP-ALS-PLS, and $\rho$-PLS as well.
On the other hand, CP-ALS-PLS and $\rho$-PLS have similar prediction accuracy and run time.
Our $\rho$-PLS method selects and smooths features so that each factor is neuroscientifically interpretable, unlike the CP-ALS-PLS approach; we demonstrate this interpretability aspect of our approach in the following section.  

\subsection{Case Study: Patient YBA}
We now show an in-depth study on patient YBA to demonstrate how $\rho$-PLS yields neuroscientifically interpretable results for brain decoding.
For ease of understanding, we group the full set of frequency bands observed in the original data together into pre-defined neuroscientific bands of interest \citep{ozker2017double}. 
Additionally, we limit our attention to looking at the sparsity in the electrodes and smoothing over time with respect to the frequency bands. 

\begin{figure}[t!]
    \centering
        \includegraphics[width=\linewidth]{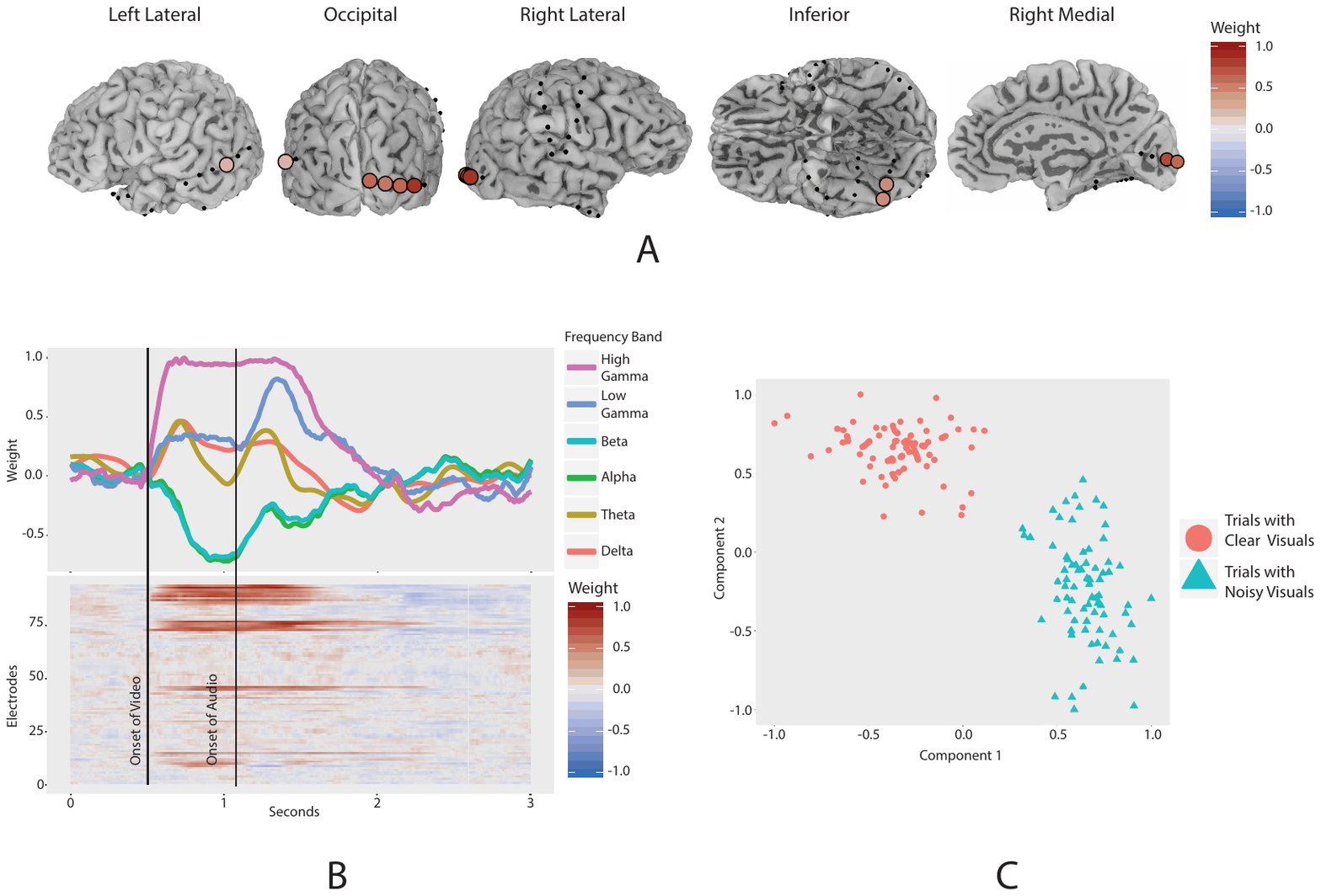}
    \caption{Visualizations of the $\rho$-PLS component for patient YBA for clear \& noisy visuals classification. 
    A.) Locations of the influential electrodes identified by $\rho$-PLS, and all are in the occipital region.
    B.) Projections of $\rho$-PLS factors communicate the behavior of the frequency bands and electrodes over time. 
    Large values in the plots correspond to large differences between signal power classes.
    C.) Projection of the $\rho$-PLS components onto the ECoG tensor. 
    The trials separate allowing for accurate classification.}
\label{fig:score}
\end{figure}

We first analyze the $\rho$-PLS factors on the noisy \& clean visual classification problem to demonstrate how $\rho$-PLS can identify important areas of interest in large ECoG tensors. 
Figure \ref{fig:score} displays interpretable visualizations of our results.

In Figure \ref{fig:score}(A), the sparse electrode factor reveals 5 electrodes in the brain which exhibit large differences in activity behaviors between the clear \& noisy visual trials. 
All of these electrodes are in the occipital region, a part of the visual cortex, which gives us confidence of the validity of our feature selection approach \citep{ozker2017double,ozker2018frontal}.
In Figure \ref{fig:score}(B), we visualize the corresponding components in the time and frequency modes in order to observe how the behaviors of the electrodes identified in Figure \ref{fig:score}(A) change over time with respect to frequency bands. 
The largest changes in response are observed immediately after the onset of the video that occurs at 0.5 seconds; this is most apparent in the High Gamma, Beta, and Alpha frequency bands.
Figure \ref{fig:score}(C) shows how the trials separate allowing for accurate classification.
It show the first two projects of the $\rho$-PLS components onto the ECoG tensor: $\mathbfcal{X} \times_2 \hat{\bm{v}}_1 \times_3 \hat{\bm{w}}_1 \times_4 \hat{\bm{t}}_1$ and $\mathbfcal{X} \times_2 \hat{\bm{v}}_2 \times_3 \hat{\bm{w}}_2 \times_4 \hat{\bm{t}}_2$.

\begin{figure}[t!]
\centering
\includegraphics[width=\linewidth]{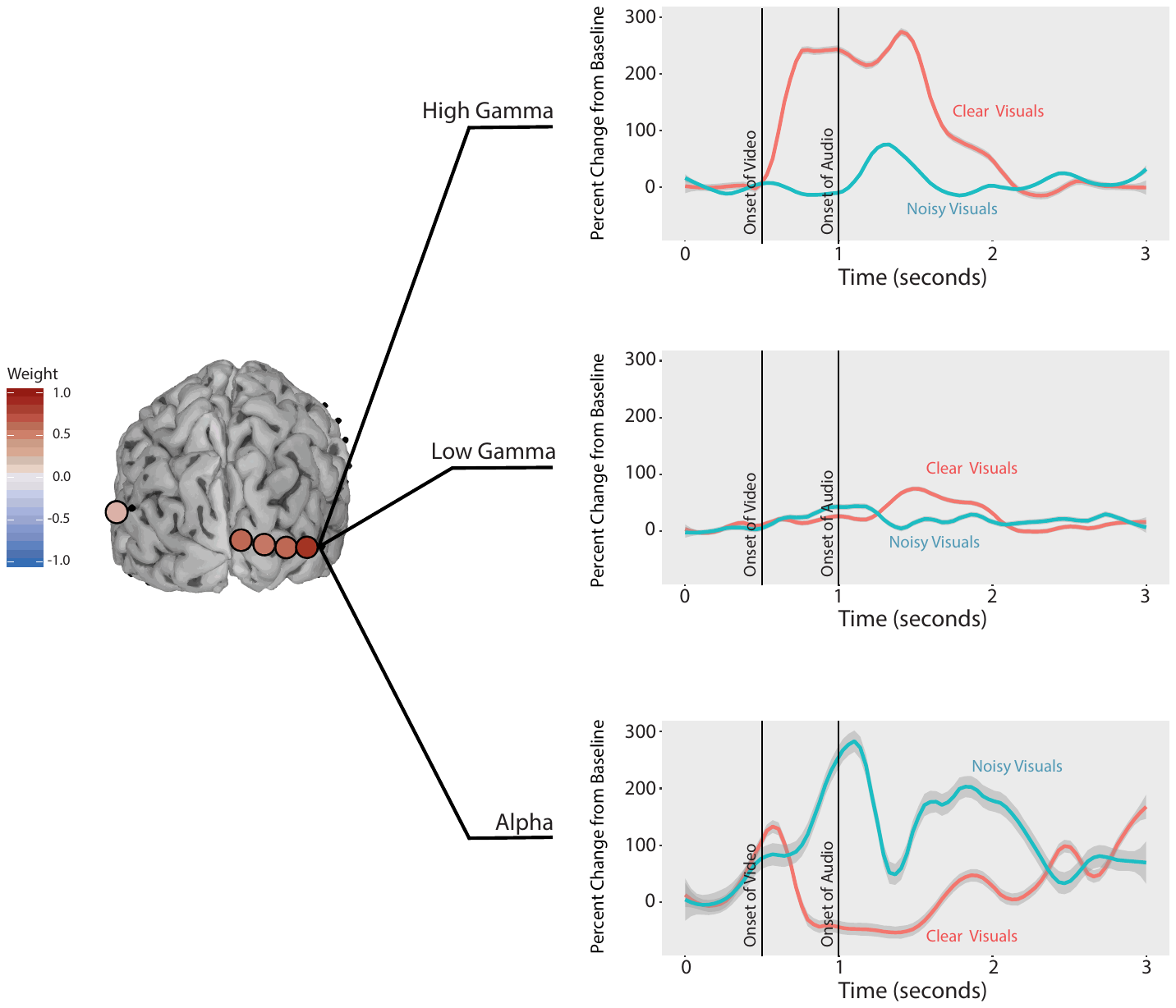}
\caption{Average percent change from baseline for patient YBA at the High Gamma, Low Gamma and Alpha frequency bands for the clear \& noisy visual classification problem. 
A locally weighted regression is fit to scaled trials. 
The grey areas around the estimate are the 95\% confidence intervals.
$\rho$-PLS identifies frequency bands, such as High Gamma and Alpha, that have a large difference in percent change from baseline.}
\label{fig:raw1}
\end{figure}

In Figure \ref{fig:raw1}, we compare the spectrograms for the High Gamma, Low Gamma, and Alpha frequency bands associated with electrode 77, one of the electrodes identified by $\rho$-PLS as influential in the noisy \& clean visual classification problem.
The activity in the High Gamma and Alpha frequency bands vary greatly. 
In particular, the activity in the High Gamma frequency band increases greatly for clear visuals, while the activity in the Alpha frequency decreases significantly.
These are common observations \citep{hipp2011oscillatory, schepers2014electrocorticography} that $\rho$-PLS is able to identify and highlight, also shown in Figure \ref{fig:score}(B).

\begin{figure}[t!]
\centering
\includegraphics[width=\linewidth]{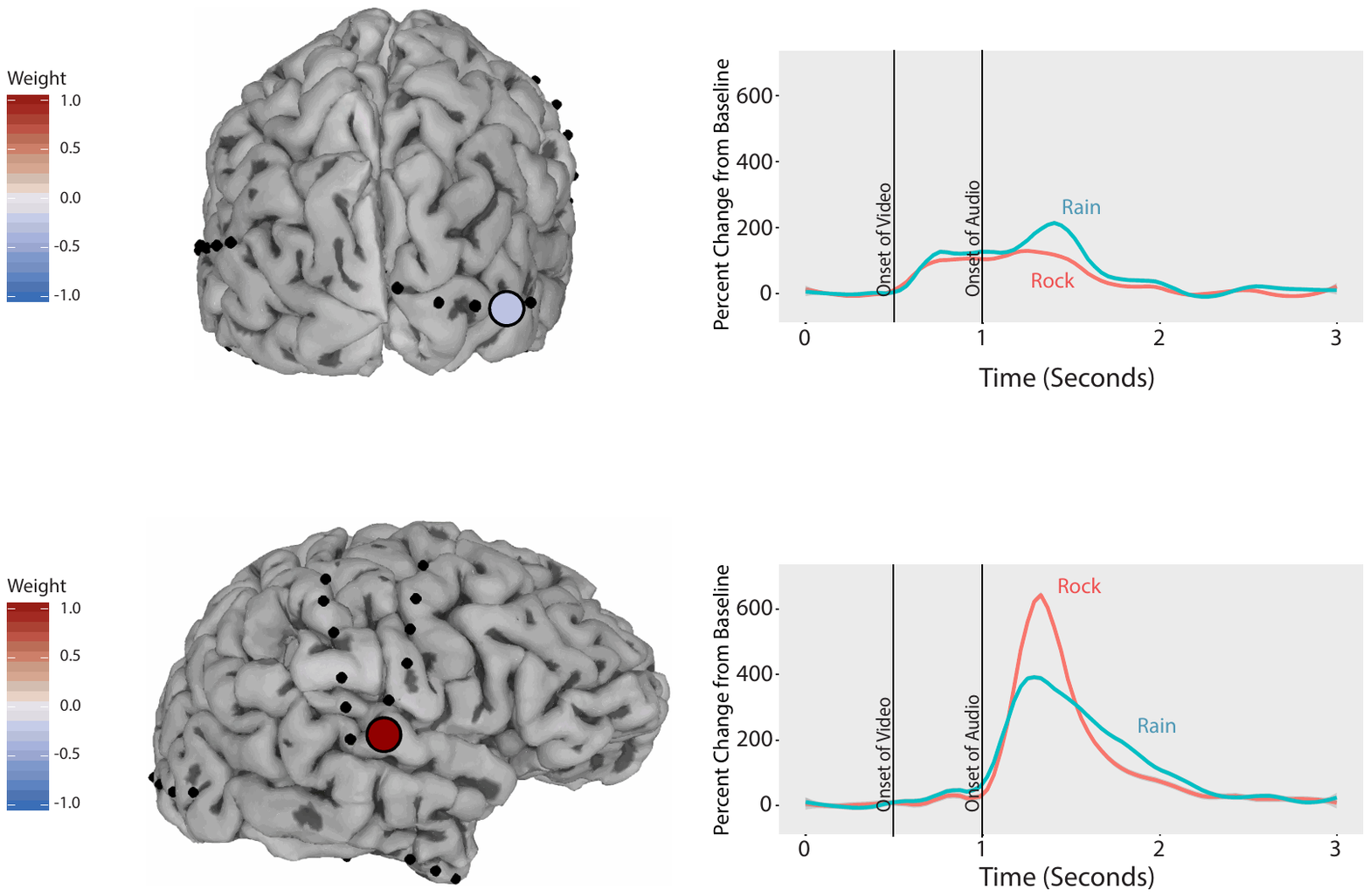}
\caption{Average percent change from baseline for patient YBA at the High Gamma frequency band for the ``rock" \& ``rain" problem. 
The magnitude of the $\rho$-PLS weight reflects the magnitude of the difference in average signal power between classes. 
Electrode 77 (top) has a small but noticeable difference in percent change from baseline, while Electrode 34 (bottom) has a large difference in percent change from baseline.
The sign of the weight indicates which class has a stronger positive response.}
\label{fig:raw2}
\end{figure}

The last example investigates how the brain responds differently to the words ``rock" and ``rain" with audio-visual speech stimuli. 
Figure \ref{fig:raw2} depicts the $\rho$-PLS electrode components for the ``rock" \&``rain" classification problem, along with their corresponding spectrograms in the High Gamma frequency band over time. 
The electrode component has two non-zero weights, one in the posterior superior temporal gyrus in the audio cortex (bottom figures), and one in the occipital region of the visual cortex (top figures).
There is a substantial difference in the High Gamma frequency band for the electrode in the posterior superior temporal gyrus; which supports previous work that connects the audio and visual cortex to speech perception \citep{stevenson2009audiovisual, ozker2017double, schepers2014electrocorticography}.

This exploratory analysis highlights existing findings, but also allows scientists to quickly investigate new findings. 
The method $\rho$-PCA is able to identify influential electrodes for predicting stimulus conditions, along with corresponding frequency bands, and our findings match what has been shown in previous work.
Thus, $\rho$-PLS offers a fast data driven method for sifting large data sets to identify regions, time points, and frequencies that warrant further study. 

\section{Discussion}
Brain decoding with ECoG data offers a unique opportunity to study the cortical processes associated with speech perception.  
However, ECoG data sets are large and complex, making them exceptionally difficult to analyze, especially for neuroscientists. 
In this paper, we introduce $\rho$-PCA and $\rho$-PLS, tensor decomposition approaches with sparse penalties and smoothing constraints that lead to directly interpretable tensor factors tailored to attributes of ECoG data. 
$\rho$-PCA is the first method we know of that considers the entirety of an ECoG data set and selects relevant features in the time points, frequencies and locations in an automated data-driven manner. 

We evaluated the classification accuracy and computational performance of both $\rho$-PCA and $\rho$-PLS and compared them to methods commonly used in neuroimaging, namely other tensor decomposition methods and classical matrix-based methods. 
During our analysis, we found that the performance of matrix methods were very dependent on implementation and some were not computationally efficient enough for us to investigate. 
On the other hand, the performance of $\rho$-PLS was on par with or better than the other tensor decomposition approaches, while also providing more interpretable solutions and visualizations.

While we have demonstrated the effectiveness of $\rho$-PCA and $\rho$-PLS specifically on ECoG data, our work can potentially be used in other neuroimaging modalities that have spatio-temporal characteristics, such as EEGs and fMRI.  
Although dimension reduction approaches have been an extremely important tool for analysis of neuroimaging data, we find that methods developed specifically for tensor data can be much more effective when the data has a natural representation as a tensor.  
Further, using constraints and regularization tailored to the attributes of the specific data set, yield more interpretable factors.  
Overall, we have developed a computationally efficient and neuroscientifically interpretable method for higher-order dimension reduction of ECoG data, with many potential uses beyond what we considered in this paper.

\section*{Supplementary Material}
The appendices of this paper can be found with the arXiv preprint, \href{https://arxiv.org/abs/2011.09447}{arXiv:2011.09447}.
The code and data are available at \url{https://github.com/DataSlingers/rho-PCA}.

\section*{Acknowledgements}
KG was partially supported by NSF award 1547433. 
AC is supported by NSF NeuroNex-1707400. 
JM and MB are supported by NIH grant R01NS065395.
GA is supported by NIH 1R01GM140468, NSF  DMS-1554821, and NSF NeuroNex-1707400.

\medskip
\bibliographystyle{apalike}
\bibliography{refs}

\clearpage

\appendix 

\subsection{Convergence of Algorithm \ref{alg::p-pca}} \label{appendix:A}

This section provides a proof for Theorem \ref{thm::bloc-sol-p-pca}, which guarantees the convergence of Algorithm \ref{alg::p-pca} to a local solution.
Proposition \ref{prop::kkt-conditions} explains the regulatory conditions for convergence.

\begin{proposition} \label{prop::kkt-conditions}
Suppose that $(\bm{u}^*, \bm{v}^*, \bm{w}^*, \bm{t}^*)$ are the global maximum of Eq. \eqref{eqn::p-pca-v2}, holding all other factors fixed.
It follows that $(\bm{u}^*, \bm{v}^*, \bm{w}^*, \bm{t}^*)$ must satisfy the Karush-Kuhn-Tucker (KKT) conditions, listed below.
The variables $\gamma_u$, $\gamma_v$, $\gamma_w$, and $\gamma_t$ are the dual variables associated with Eq. \eqref{eqn::p-pca-v2}.
The notation $\tilde{\nabla}_x f(\bm{x})$ represents a general sub-gradient of $f$ with respect to $x$.
\vspace{0pt}
\begin{equation*}
\begin{split}
     \left(\mathbfcal{X} \times_2 \bm{v} \times_3 \bm{w} \times_4 \bm{t} \right) & \\ - 2 \gamma_u \bm{u}^* &= 0 \quad (\bm{u}\text{-stationarity}) \\
     \gamma_u \left(\norm{\bm{u}^*}_2^2 - 1\right) &= 0 \quad (\bm{u}\text{-complementary} \\ &  \qquad \quad \text{slackness})\\
     \gamma_u &\geq 0 \quad (\bm{u}\text{-dual feasibility})\\
     \norm{\bm{u}^*}_2^2 &\leq 1 \quad (\bm{u}\text{-primal feasibility})\\
     \left(\mathbfcal{X} \times_1 \bm{u} \times_3 \bm{w} \times_4 \bm{t} \right) & \\ - \lambda_v \tilde{\nabla}_v \left(\norm{\bm{v}^*}_1\right)  - 2 \gamma_v \bm{v}^* &= 0 \quad (\bm{v}\text{-stationarity}) \\     
     \gamma_v \left(\norm{\bm{v}^*}_2^2 - 1\right) &= 0 \quad (\bm{v}\text{-complementary} \\ &  \qquad \quad \text{slackness})\\
     \gamma_v &\geq 0 \quad (\bm{v}\text{-dual feasibility})\\
     \norm{\bm{v}^*}_2^2 &\leq 1 \quad (\bm{v}\text{-primal feasibility})\\
     \left(\mathbfcal{X} \times_1 \bm{u} \times_2 \bm{v} \times_4 \bm{t} \right) & \\ - \lambda_w \tilde{\nabla}_w \left(\norm{\bm{w}^*}_1\right)  - 2 \gamma_w \bm{S}_w &= 0 \quad (\bm{w}\text{-stationarity}) \\
     \gamma_w \left(\norm{\bm{w}^*}_{\bm{S}_w}^2 - 1\right) &= 0 \quad (\bm{w}\text{-complementary} \\ &  \qquad \quad \text{slackness})\\
     \gamma_w &\geq 0 \quad (\bm{w}\text{-dual feasibility}) \\
     \norm{\bm{w}^*}_{\bm{S}_w}^2 &\leq 1 \quad (\bm{w}\text{-primal feasibility}) \\
     \left(\mathbfcal{X} \times_1 \bm{u} \times_2 \bm{v} \times_3 \bm{w} \right) & \\- 2 \gamma_t \bm{S}_t \bm{t}^* &= 0 \quad (\bm{t}\text{-stationarity})\\
     \gamma_t \left(\norm{\bm{t}^*}_{\bm{S}_t}^2 - 1\right) &= 0 \quad (\bm{t}\text{-complementary} \\ &  \qquad \quad \text{slackness})\\
     \gamma_t &\geq 0 \quad (\bm{t}\text{-dual feasibility})\\
     \norm{\bm{t}^*}_{\bm{S}_t}^2 &\leq 1 \quad (\bm{t}\text{-primal feasibility})
\end{split}
\end{equation*}
\end{proposition}
 
\medskip
 
\noindent
\textit{Proof of Theorem \ref{thm::bloc-sol-p-pca}}:
We follow the approach used by \citet{allen2013multi}, and rely on results from \citet{tseng2001convergence}.
\citet{allen2011sparse} states that we can solve for one factor at a time, holding the others fixed at feasible values.
In order for this to hold, we show that each factor is a unique global solution of their respective regression problems.
Additionally, we demonstrate that each factor is a local solution to Eq. \eqref{eqn::p-pca-v2}, holding all others fixed.
Lastly, we will derive the updates for each factor, shown in Equations \eqref{eqn::u-update}-\eqref{eqn::t-update} of Alg. \ref{alg::p-pca}.

Notice that $\mathbfcal{X} \times_1 \bm{u} \times_2 \bm{v} \times_3 \bm{w} \times_4 \bm{t}$, is continuously differentiable with respect to all factors.
Each of the optimization problems for $\bm{u}$, $\bm{v}$, $\bm{w}$ and $\bm{t}$ are strictly concave.
Alg. \ref{alg::p-pca} requires that $\bm{\Omega} \succeq 0$ so that the matrix $I + \alpha_t\bm{\Omega}_t$ is positive definite by construction. 
Computing the Hessian of the Lagrangian with respect to $\bm{t}$ yields $- (\bm{I} + \alpha_t\bm{\Omega}_t)$, a negative definite matrix making the problem strictly concave.
This implies that the solution $\bm{t}^*$ is unique. 
If we consider the Lagrangian for the other vectors, we can show that the respective optimization problems are strictly concave. 
The Lagrangian of the optimization problem with respect to $\bm{v}$, where we take the square root of the constraint, is,
\vspace{0pt}
\begin{equation*}
    \begin{aligned}
        \underset{\bm{v}}{\text{maximize}} &\quad \mathbfcal{X} \times_1  \bm{u} \times_2  \bm{v} \times_3  \bm{w} \times_4  \bm{t} \\
        &- \lambda \norm{\bm{v}}_1 -  \gamma(\norm{\bm{v}}_2^2 - 1)
    \end{aligned}
\end{equation*}
\vspace{0pt}
Using the Triangle Inequality and the definition of concavity, it follows that,
\begin{equation*}
\begin{aligned}
    \mathbfcal{X} \times_1 \bm{u} \times_2 (c\bm{x} + (1-c)\bm{y}) \times_3 \bm{w} \times_4 \bm{t} \\
        - \lambda_v \norm{c\bm{x} + (1-c)\bm{y}}_1 - (\norm{c\bm{x} + (1-c)\bm{y}}_2^2 - 1) \\
        > c\left(\mathbfcal{X} \times_1 \bm{u} \times_2 \bm{x} \times_3 \bm{w} \times_4 \bm{t}\right) \\ 
        - c \lambda_v \norm{\bm{x}}_1 - c(\norm{\bm{x}}_2^2 - 1) \\
         + (1-c)\left(\mathbfcal{X} \times_1 \bm{u} \times_2 \bm{y} \times_3 \bm{w} \times_4 \bm{t}\right) \\
         - (1-c) \lambda_v \norm{\bm{y}}_1 - (1-c)(\norm{\bm{y}}_2^2 - 1),
\end{aligned}
\end{equation*}
where the problem is strictly concave, which makes $\bm{v}^*$ unique. 
Similar arguments hold for $\bm{u}$ \citep{allen2012sparse, allen2013multi} and $\bm{w}$ \citep{Allen:2019}. 
Together these properties imply that our problem converges to a stationary point of the joint optimization problem. 
Next, we will derive the block coordinate updates for Alg. \ref{alg::p-pca}.
\begin{enumerate}[(i.)]\itemsep0pt 
    \item \textbf{Derive the coordinate block solution for the trial mode.}
    First, we solve $\bm{u}$-stationarity to find the optimal $\bm{u}^*$,
     \vspace{0pt}
     \begin{align*}
         \bm{u}^* = \frac{1}{2 \gamma_u} \left(\mathbfcal{X} \times_2 \bm{v} \times_3 \bm{w} \times_4 \bm{t} \right) 
     \end{align*}
     Define $\tilde{\bm{u}} = \mathbfcal{X} \times_2 \bm{v} \times_3 \bm{w} \times_4 \bm{t}$.
     We will use this result with $\bm{u}$-complementary slackness to find $\gamma_u^*$,
     \begin{equation*}
     \begin{split}
         \gamma_u \left(\norm{\bm{u}^*}_2^2 - 1\right) = \gamma_u \left(\frac{1}{4 \gamma_u^2} \tilde{\bm{u}}^T \tilde{\bm{u}} - 1\right) = 0  \\
     \end{split}
     \end{equation*}
     \begin{align*}
         \gamma_u^{*2} &= \frac{1}{4} \tilde{\bm{u}}^T \tilde{\bm{u}} = \frac{1}{4}\norm{\tilde{\bm{u}}}_2^2
     \end{align*}
     Plugging $\gamma_u^*$ into $\bm{u}^*$ yields the coordinate update Eq. \eqref{eqn::u-update}.
     By Proposition \ref{prop::kkt-conditions} and Proposition 1 from \citet{allen2012sparse}, $\bm{u}^*$ is a global optimum of its sub-problem, and local optimum of Eq. \eqref{eqn::p-pca-v2} holding all other factors fixed.
    
     \item \textbf{Derive the coordinate block solution for the electrode mode.}
     Our approach to finding the optimal $(\bm{v}^*, \gamma_v^*)$ follows those by \citet{allen2012sparse} and \citet{witten2009penalized}.
     First, we will solve $\bm{v}$-stationarity to find the optimal $\bm{v}^*$.
    Define $\tilde{\bm{v}} = \mathbfcal{X} \times_1 \bm{u} \times_3 \bm{w} \times_4 \bm{t}$.
     We can determine the value of each $v_i^*$ element of $\bm{v}^*$ as follows,
     \begin{align*}
           2 \gamma_v v_i^* =
           \begin{cases}
             \left(\tilde{\bm{v}}\right)_i - \lambda_v & v_i > 0, \\
             \left(\tilde{\bm{v}}\right)_i + \lambda_v & v_i < 0, \\
             \left(\tilde{\bm{v}}\right)_i - [-\lambda_v, \lambda_v], & v_i = 0 
         \end{cases} 
     \end{align*}
     We can alternatively express $\bm{v}^*$ as a function of the soft-thresholding function,
     \begin{align*}
         \bm{v}^* &= \frac{1}{2 \gamma_v} S(\tilde{\bm{v}}, \lambda_v) 
     \end{align*}
     Now, we can use $\bm{v}$-complementary slackness to find $\gamma_v^*$,
     \begin{equation*}
     \begin{split}
         \gamma_v \left(\norm{\bm{v}^*}_2^2 - 1\right) & = \gamma_v \left(\frac{1}{4 \gamma_v^2} S(\tilde{\bm{v}}, \lambda_v)^T S(\tilde{\bm{v}}, \lambda_v) - 1\right) \\
         &= 0 
     \end{split}
     \end{equation*}
     \begin{equation*}
     \begin{split}
         \gamma_v^{*2} &= \frac{1}{4} S(\tilde{\bm{v}}, \lambda_v)^T  S(\tilde{\bm{v}}, \lambda_v) = \frac{1}{4}\norm{S(\tilde{\bm{v}}, \lambda_v)}_2^2 
     \end{split}
     \end{equation*}
     Plugging $\gamma_v^*$ into $\bm{v}^*$ results in the coordinate update Eq. \eqref{eqn::v-update}.
     By Proposition \ref{prop::kkt-conditions} and Theorem 1 of \citet{allen2012sparse}, $\bm{v}^*$ is a global optimum of its sub-problem, and local optimum of Eq. \eqref{eqn::p-pca-v2}, holding all other factors fixed.

     \item \textbf{Derive the coordinate block solution for the frequency mode.}
     Our approach to finding the optimal $(\bm{w}^*, \gamma_w^*)$ follows those from \citet{Allen:2019} and \citet{allen2014generalized}.
     Notice that,
     \begin{equation} \label{eqn::prox-1}
     \begin{split}
         \underset{\bm{w} \in \mathbb{R}^q}{\mathrm{argmin}} & \; \underbrace{\frac{\bm{w}^T \bm{S}_w \bm{w}}{2} - \bm{w}^T \left(\mathbfcal{X} \times_1 \bm{u} \times_2 \bm{v} \times_4 \bm{t} \right)}_{\text{smooth}} \\ & \quad + \underbrace{\lambda_w \norm{\bm{w}}_1 + i_{\bar{B}^{q}_{\bm{S}_w}}(\bm{w})}_{\text{non-differentiable}} 
         \end{split}
     \end{equation}
     \vspace{0pt}
     We define $\bar{\mathbb{B}}_{\bm{S}_w}^{q}$ as the unit ellipse of the $\bm{S}_w$-norm, $\bar{\mathbb{B}}_{\bm{S}_w}^{q} = \{\bm{w} \in \mathbb{R}^q | \bm{w}^T \bm{S}_w \bm{w} \leq 1\}$.
     The term $i_{\bar{\mathbb{B}}_{\bm{S}_w}^{q}}$ is the so called infinite indicator of the feasible set of $\bm{w}$;
     defined by
     \begin{equation*} 
         i_{\mathcal{X}}(x) = \begin{cases}
             0, & \text{$x$ is an element of $\mathcal{X}$}, \\
             +\infty, & \text{otherwise}
 	\end{cases} 
     \end{equation*}
     Due to the fact that Eq. \eqref{eqn::prox-1} is a separable sum of convex functions, we can use the proximal operator to find the coordinate update for $\bm{w}^*$ \citep{parikh2014proximal}.
     The proximal operator is defined as,
     \begin{equation*}
         \text{prox}_{\lambda f}(\bm{y}) = \argmin_{\bm{x}} \left(f(\bm{x}) + \frac{1}{2 \lambda} \norm{\bm{x} - \bm{y}}_2^2\right),
     \end{equation*}
     where $\lambda > 0$ is a constant and $f$ is a function.
     In the context of Eq. \eqref{eqn::prox-1}, we take $f(\bm{x})= \norm{\bm{x}}_1$ and find that, 
     \begin{equation*}
         \begin{split}
             \text{prox}_{f + i_{\bar{\mathbb{B}}_{\bm{S}_w}^{q}}}(\bm{x}) &= \argmin_{\bm{w}} f(\bm{x}) + \frac{1}{2} \norm{\bm{x} - \bm{w}}_2^2  \\ & \quad  + i_{\bar{\mathbb{B}}_{\bm{S}_w}^{q}}(\bm{w}) \\
             &= \argmin_{\bm{w} \in i_{\bar{\mathbb{B}}_{\bm{S}_w}^{q}}} f(\bm{x}) + \frac{1}{2} \norm{\bm{x} - \bm{w}}_2^2  \\
             &= \text{proj}_{i_{\bar{\mathbb{B}}_{\bm{S}_w}^{q}}} \left(\text{prox}_f (\bm{x})\right)
         \end{split}
     \end{equation*}
     where $\text{proj}_{\mathcal{X}}(\bm{x})$ denotes the projection of $\bm{x}$ onto $\mathcal{X}$.
     This result immediately follows from \citet{Allen:2019} and \citet{yu2013decomposing}; Eq. \eqref{eqn::w-update} is the coordinate update for the frequency mode.
     By Proposition \ref{prop::kkt-conditions} and Lemma 3 of \citet{Allen:2019}, $\bm{w}^*$ is a global optimum of its sub-problem, and local optimum of Eq. \eqref{eqn::p-pca-v2}, holding all other factors fixed.
     
     \item \textbf{Derive the coordinate block solution for the time mode.}
     Our approach to finding the optimal $(\bm{t}^*, \gamma_t^*)$ that from \citet{allen2013multi}.
    Define $\tilde{\bm{t}} = \mathbfcal{X} \times_1 \bm{u} \times_2 \bm{v} \times_3 \bm{w}$.
     We will solve $\bm{t}$-stationarity to find the optimal $\bm{t}^*$, 
     \begin{equation*}
         \bm{t}^* =\frac{1}{2 \gamma_t} \bm{S}_t^{-1} \tilde{\bm{t}} 
     \end{equation*}
     Next, we will use this results with $\bm{t}$-complementary slackness to find $\gamma_t^*$:
     \begin{equation*}
     \begin{split}
         0 &= \gamma_t \left(\norm{\bm{t}^*}_{\bm{S}_t}^2 - 1\right) \\
             &= \gamma_t \left(\frac{1}{4 \gamma_t^2}\tilde{\bm{t}}^T \bm{S}_t^{-T} \bm{S}_t \bm{S}_t^{-1}\tilde{\bm{t}} - 1\right) \\
             &= \gamma_t \left(\frac{1}{4 \gamma_t^2}\tilde{\bm{t}}^T \bm{S}_t^{-1}\tilde{\bm{t}} - 1\right) \\
             &= \gamma_t \left(\frac{1}{4 \gamma_t^2} \norm{ \tilde{\bm{t}}}_{\bm{S}_t^{-1}}^2 - 1\right) 
     \end{split}
     \end{equation*}
     \begin{align*}
         \gamma_t^{*2} = \frac{1}{4} \norm{ \tilde{\bm{t}} }_{\bm{S}_t^{-1}}^2
     \end{align*}
     Plugging $\gamma_t^*$ into $\bm{t}^*$ yields the coordinate update Eq. \eqref{eqn::t-update}.
     By Proposition \ref{prop::kkt-conditions}, $\bm{t}^*$ is a global optimum of its sub-problem, and local optimum of Eq. \eqref{eqn::p-pca-v2}, holding all other factors fixed.
 \end{enumerate}
We conclude that $\rho$-PCA converges to a local optimum, as well as find the block coordinate updates. 
$\square$

\subsection{Model Selection: Tuning $\rho$-PCA and $\rho$-PLS}

\subsubsection{Tuning $\rho$-PCA: Nested Bayesian Information Criterion.}
The performance of $\rho$-PCA can be improved by selecting values of parameters that yield the best model fit criteria. 
There are several unknown parameters in Algorithm \ref{alg::p-pca}: $\lambda_v$, $\lambda_w$, $\alpha_w$, and $\alpha_t$.  
We suggest the tuning $\rho$-PCA using the nested Bayesian Information Criterion (BIC) schema \citep{lee2010biclustering, allen2011sparse}. 
The BIC identifies the most promising candidate parameters based, in part, on the likelihood of the model \citep{schwarz1978estimating}. 
In the case of the Algorithm \ref{alg::p-pca}, we cannot base our BIC on the likelihood, as the penalty selection is different for each mode. 
In response, Proposition \ref{prop::bic} defines the BIC for each mode separately, based on their respective regression problem. 

\begin{proposition} \label{prop::bic}
We can calculate BIC for each mode as follows,
\begin{equation} \label{eqn::v-bic}
\begin{aligned}
    \text{BIC}(\tilde{\bm{v}})  = & \log \left(\frac{1}{p} \norm{\left(\mathbfcal{X} \times_1 \hat{\bm{u}} \times_3 \hat{\bm{w}} \times_4 \hat{\bm{t}}\right) - \tilde{\bm{v}}}_2^2 \right) \\ & + \frac{1}{p} \log(p) \cdot \norm{\tilde{\bm{v}}}_0,
\end{aligned}
\end{equation}
\begin{equation} \label{eqn::w-bic}
\begin{aligned}
    \text{BIC}(\tilde{\bm{w}}) = & \log \left(\frac{1}{q} \norm{\left(\mathbfcal{X} \times_1 \hat{\bm{u}} \times_2 \hat{\bm{v}} \times_4 \hat{\bm{t}}\right) - \tilde{\bm{w}}}_2^2 \right) \\ & + \frac{1}{q} \log(q) \cdot \text{Tr}\left[\left(\bm{I}_{\abs{\mathcal{A}}} + \alpha_w \bm{\Omega}_w^{\mathcal{A}}\right)^{-1}\right],
\end{aligned}
\end{equation}
\noindent where $\mathcal{A}$ represents the indices of the active set, or estimated non-zero elements of $\hat{\bm{w}}$, and $\bm{\Omega}_w^{\mathcal{A}}$ is the corresponding submatrix of $\bm{\Omega}_w$,
\begin{equation} \label{eqn::t-bic}
\begin{aligned}
    \text{BIC}(\tilde{\bm{t}}) = & \log \left(\frac{1}{r} \norm{\left(\mathbfcal{X} \times_1 \hat{\bm{u}} \times_2 \hat{\bm{v}} \times_3 \hat{\bm{w}} \right) - \tilde{\bm{t}}}_2^2 \right) \\ & + \frac{1}{r} \log(r) \cdot \text{Tr}\left[\bm{S}_t\right]
\end{aligned}
\end{equation}
\end{proposition}

Notice that the BIC formulations \eqref{eqn::v-bic}-\eqref{eqn::t-bic} are of similar form, with the exception of degrees of freedom in the second additive term.
Each penalty we use in $\rho$-PCA is has a different calculation for its associated degrees of freedom.
The degrees of freedom for electrodes ($\bm{v}$), frequency ($\bm{w}$), and time ($\bm{t}$) are derived from \citet{allen2013multi}, \citet{Allen:2019}, and \citet{goutte2010probabilistic}, respectively.
The mode trial ($\bm{u}$) is excluded, since it has no associated parameter.
Algorithm \ref{alg::nested-bic} describes model selection for $\rho$-PCA in detail.

\subsubsection{Tuning $\rho$-PLS: Nested Cross-validation.}
Just like $\rho$-PCA, $\rho$-PLS has unknown hyperparameters $\lambda_v$, $\lambda_w$, $\alpha_w$, and $\alpha_t$.
To select an optimal $\rho$-PLS model given the data at hand, we implement a nested $N$-fold cross-validation scheme within each factor $K$.
This approach provides control over the regularization in each component. 
For each factor $k = 1, \ldots, K$, we select the set of regularization parameters that maximize prediction accuracy for all of the currently computed components. 
We cache the results of each run so that we can deflate the covariance tensor appropriately.
This procedure described in Algorithm \ref{alg::nested-cv}. 

\subsection{Cumulative Amount of Variance Explained}
Principal Components Analysis computes a set of new variables called \textit{Principal Components (PCs)} or \textit{factor loadings}.
The PCs describe the amount of variance explained by the covariates.
Similarly, tensor PCs can be computed from a tensor decompositions \citep{allen2012sparse}.
In the context of $\rho$-PCA, the estimated $K$ component vectors for each mode are: 
$$\hat{\bm{U}} := \{\hat{\bm{u}}_k\}_{k=1}^{K}, \hat{\bm{V}} := \{\hat{\bm{v}}\}_{k=1}^{K},$$ $$\hat{\bm{W}}_k := \{\hat{\bm{w}}_k\}_{k=1}^{K}, \text{ and } \hat{\bm{T}}_k := \{\hat{\bm{t}}_k\}_{k=1}^{K}$$
In turn, these projections can be used to find the cumulative proportion of variance explained and evaluate the estimation of $\rho$-PCA, noted by Proposition \ref{prop::ave} \citep{allen2011sparse}:

\begin{proposition} [\citet{allen2011sparse}] \label{prop::ave}
Define $\hat{\bm{U}}_k = \{\hat{\bm{u}}_i\}_{i=1}^{k} \in \mathbb{R}^{n \times k}$, and the projection matrix $\bm{P}_k^{(U)} = \hat{\bm{U}}_k \left(\hat{\bm{U}}_k^T \hat{\bm{U}}_k\right)^{-1} \hat{\bm{U}}_k^T$.
The matrices $\bm{P}_k^{(V)}$, $\bm{P}_k^{(W)}$, and $\bm{P}_k^{(T)}$ can be defined analogously.
The cumulative amount of variance explained (CAVE) by the first $k$ higher-order principal components is given by 
\begin{equation*} \label{eqn::ave}
    \text{CAVE}_k = \frac{\norm{\mathbfcal{X} \times_1 \bm{P}_k^{(U)} \times_2 \bm{P}_k^{(V)} \times_3 \bm{P}_k^{(W)} \times_4 \bm{P}_k^{(T)}}^2_F}{\norm{\mathbfcal{X}}^2_F}
\end{equation*}
\end{proposition}

\begin{algorithm*}
\caption{Model selection for $\rho$-PCA: Nested BIC}\label{alg::nested-bic}
\begin{enumerate}\itemsep0pt 

    \item Initialization:
    \begin{enumerate}[(A)]\itemsep0pt 
        \item Set the maximum iterations $B$, $\hat{\mathbfcal{X}} = \mathbfcal{X}$, candidate parameters $\{\lambda_v\}_{i=1}^{P_v}$, $\{\left(\lambda_w, \alpha_w\right)\}_{i=1}^{P_w}$, and $\{\alpha_t\}_{i=1}^{P_t}$.
        \item Initialize $\left\{\left(\hat{d}_k, \hat{\bm{u}}_k, \hat{\bm{v}}_k, \hat{\bm{w}}_k, \hat{\bm{t}}_k\right)\right\}_{k=1}^{K}$ using CP-decomposition.
    \end{enumerate}
    
    \item For $k = 1, \ldots, K$:
    \begin{enumerate}[(A)]\itemsep0pt 
    
        \item Repeat until convergence, or for a maximum of $B$ iterations:
        \begin{enumerate}[i.]\itemsep0pt 
        
            \item Update $\hat{\bm{u}}_k$ using Eq. \eqref{eqn::u-update} from Algorithm \ref{alg::p-pca}.
            
            \item Update $\hat{\bm{v}}_k$,
            \begin{enumerate}[(a)]\itemsep0pt 
                
                \item For each candidate parameter $\lambda_{v_i}$, $i = 1, \ldots P_v$.
                \begin{itemize}
                    
                    \item[$-$] Calculate, $\Breve{\bm{v}}_k(\lambda_{v_i}) = S(\hat{\mathbfcal{X}} \times_1 \hat{\bm{u}}_{k-1} \times_3 \hat{\bm{w}}_{k-1} \times_4 \hat{\bm{t}}_{k-1},\lambda_{v_i})$
                    
                    \item[$-$] Find the BIC using $\Breve{\bm{v}}_k(\lambda_{v_i})$ and Eq. \eqref{eqn::v-bic}, denoted as $\text{BIC}(\Breve{\bm{v}}_k(\lambda_{v_i}))$.
                    
                    \item[$-$] Select the optimal parameter $\lambda_{v_i}^*$ and factor, $\tilde{\bm{v}}_k = \argmin_{\lambda_{v_i}} \text{BIC}(\Breve{\bm{v}}_k(\lambda_{v_i}))$.
                    
                    \item[$-$] Re-scale the factor to obtain $\hat{\bm{v}}_k$, $                        \hat{\bm{v}}_k = 
                        \begin{cases}
                		\frac{\tilde{\bm{v}}_k}{\norm{\tilde{\bm{v}}_k}_2}, & \norm{\tilde{\bm{v}}_k}_2 > 0, \\
                		0, & otherwise
                	\end{cases}$
                \end{itemize}
            \end{enumerate}
            
            \item Update $\hat{\bm{w}}_k$ using a single proximal update
            \begin{enumerate}[(a)]\itemsep0pt 
            
                \item For each set of candidate parameters $\left(\lambda_w, \alpha_w\right)_i$, $i = 1, \ldots, P_w$,
                \begin{itemize}
                
                    \item[$-$] Calculate $\Breve{\bm{w}}_k(\lambda_{w_i}, \alpha_{w_i})$, $\bm{S}_{w_i} = \bm{I}_q + \alpha_{w_i}\bm{\Omega}_w$, and
                    \vspace{1pt}
                    \begin{equation*}
                    \begin{split}
                        \Breve{\bm{w}}_k(\lambda_{w_i}, \alpha_{w_i}) &= S \left(\hat{\bm{w}}_{k-1} + L_{w_i}^{-1} \left( \hat{\mathbfcal{X}} \times_1 \hat{\bm{u}}_{k-1} \times_2 \hat{\bm{v}}_{k-1} \times_4 \hat{\bm{t}}_{k-1}   - \bm{S}_{w_i} \hat{\bm{w}}_{k-1}\right), \frac{\lambda_{w_i}}{L_{w_i}} \right)
                    \end{split}
                    \end{equation*}
                    
                    \item[$-$] Find the BIC using $\Breve{\bm{w}}_k(\lambda_{w_i}, \alpha_{w_i})$ and Eq. \eqref{eqn::w-bic}, denoted as $\text{BIC}(\Breve{\bm{w}}_k(\lambda_{w_i}, \alpha_{w_i}))$.
                    
                    \item[$-$] Select the optimal parameters $(\lambda_{w_i}^*, \alpha_{w_i}^*)$, and factor, 
                    \vspace{1pt}
                    \begin{equation*}
                        \tilde{\bm{w}}_k = \argmin_{(\lambda_{w_i}, \alpha_{w_i})} \text{BIC}(\Breve{\bm{w}}_k(\lambda_{w_i}, \alpha_{w_i}))
                    \end{equation*}
                    
                    \item[$-$] Re-scale the factor to get $\hat{\bm{w}}_k$, $                        \hat{\bm{w}} = 
                      	\begin{cases}	\frac{\tilde{\bm{w}}_k}{\norm{\tilde{\bm{w}}_k}_{\bm{S}_{w_i}^{*}}}, & \norm{\tilde{\bm{w}}_k}_{\bm{S}_{w_i}^{*}} > 0, \\
                    		0, & otherwise		
                    	\end{cases}$
                \end{itemize}
            \end{enumerate}

            \item Update $\hat{\bm{t}}_k$,
            \begin{enumerate}[(a)]\itemsep0pt 
                \item For each parameter $\alpha_{t_i}$, $i = 1, \ldots, P_t$,
                \begin{itemize}
                    \item [$-$] Calculate $\Breve{\bm{t}}_k(\alpha_{t_i})$, $\bm{S}_{t_i} = \bm{I}_r + \alpha_{t_i}\bm{\Omega}_t$, and $\Breve{\bm{t}}_k(\alpha_{t_i}) = \bm{S}_{t}^{-1} \left( \hat{\mathbfcal{X}} \times_1 \hat{\bm{u}}_{k-1} \times_2 \hat{\bm{v}}_{k-1} \times_3 \hat{\bm{w}}_{k-1} \right)$.
                    \item[$-$] Find the BIC using $\Breve{\bm{t}}_k(\alpha_{t_i})$ and Eq. \eqref{eqn::t-bic}, denoted as $\text{BIC}(\Breve{\bm{t}}_k(\alpha_{t_i}))$
                    \item[$-$] Select the optimal parameter $\alpha_{t_i}$, and factor $\tilde{\bm{t}}_k = \argmin_{\alpha_{t_i}} \text{BIC}(\Breve{\bm{t}}_k(\alpha_{t_i}))$.
                    \item[$-$] Re-scale the factor for $\hat{\bm{t}}_k$, $                        \hat{\bm{t}} = 
                      	\begin{cases}	\frac{\tilde{\bm{t}}_k}{\norm{\tilde{\bm{t}}_k}_{\bm{S}_{t_i}^{*}}}, & \norm{\tilde{\bm{t}}_k}_{\bm{S}_{t_i}^{*}} > 0, \\
                    		0, & otherwise		
                    	\end{cases}$
                \end{itemize}
                \end{enumerate}
            \end{enumerate}
        
        \item Estimate $\hat{d}_k$ and deflate $\hat{\mathbfcal{X}}$,
        \begin{equation*}
        \begin{split}
            \hat{d}_k &= \hat{\mathbfcal{X}} \times_1 \hat{\bm{u}}_k \times_2 \hat{\bm{v}}_k \times_3 \hat{\bm{w}}_k \times_4 \hat{\bm{t}}_k, \\
            \hat{\mathbfcal{X}} &= \hat{\mathbfcal{X}} - \hat{d}_k \cdot \hat{\bm{u}}_k \circ \hat{\bm{v}}_k \circ \hat{\bm{w}}_k \circ \hat{\bm{t}}_k
        \end{split}
        \end{equation*}
        \item Cache the optimal tuning parameters $(\lambda_v^*, \lambda_w^*, \alpha_w^*, \alpha_t^*)$ for $k$.
    \end{enumerate}
    \item Return the optimal parameters $\{(\lambda_v^*, \lambda_w^*, \alpha_w^*, \alpha_t^*)_k\}_{k=1}^{K}$.
\end{enumerate}
\end{algorithm*}

\begin{algorithm*} 
\caption{Model selection for $\rho$-PLS: Nested $N$-fold cross validation} \label{alg::nested-cv}
\begin{enumerate}\itemsep0pt 
    \item Initialization:
    \begin{enumerate}[(A)]\itemsep0pt 
        \item Define the number of folds $B$, and sets of candidate parameters $\bm{\lambda} = \left\{\left(\lambda_v, \lambda_w, \alpha_w, \alpha_t\right)_j\right\}_{j=1}^{J}$.
        \item Split data into $B$ parts to create training $\left(\mathbfcal{X}_b^{tr}, \bm{y}_b^{tr}\right)$ and testing $\left(\mathbfcal{X}_b^{ts}, \bm{y}_b^{ts}\right)$ sets. Center all responses to obtain $\bar{\bm{y}}_b^{tr}$ and $\bar{\bm{y}}_b^{ts}$ for each fold.
    \end{enumerate}
    
    \item For each component $k = 1, \ldots, K$,
    \begin{enumerate}[(A)]\itemsep0pt 
    
        \item For each fold $b = 1, \ldots, B$,
        \begin{enumerate}[i.]\itemsep0pt 
        
            \item For each set of candidate parameters $\bm{\lambda}_j = \left(\lambda_v, \lambda_w, \alpha_w, \alpha_t\right)_j$, $j = 1, \ldots, J$,
            \begin{enumerate}[(a)]\itemsep0pt 
                \item Compute the covariance tensor from training data, $\mathbfcal{Z}_b = \mathbfcal{X}_b \times_1 \bar{\bm{y}}_b^{tr}$, and set $\hat{\mathbfcal{Z}}_b = \mathbfcal{Z}_b$.

                \item Using the $\rho$-PCA step for component $k$ from Algorithm \ref{alg::p-pca}, compute the components of $\hat{\mathbfcal{Z}}_b$,
                \vspace{1pt}
                \begin{equation*}
                    \left[\hat{\bm{v}}_{jbk}, \hat{\bm{w}}_{jbk}, \hat{\bm{t}}_{jbk} \right] = \rho \text{-PCA}(\hat{\mathbfcal{Z}}_b, k, \lambda_j)
                \end{equation*}
                
                \item Reduce the dimensionality of the data,
                \vspace{1pt}
                \begin{equation*}
                    \begin{split}
                        \Tilde{\bm{X}}_{jbk}^{tr} &= \mathbfcal{X}_b^{tr} \times_2 \hat{\bm{v}}_{jbk} \times_3 \hat{\bm{w}}_{jbk} \times_4 \hat{\bm{t}}_{jbk}  \\
                        \Tilde{\bm{X}}_{jbk}^{ts} &= \mathbfcal{X}_b^{ts} \times_2 \hat{\bm{v}}_{jbk} \times_3 \hat{\bm{w}}_{jbk} \times_4 \hat{\bm{t}}_{jbk} \\
                    \end{split}
                \end{equation*}
                
                \item Predict the class memberships of $\bm{y}_b^{ts}$ using choice SLM, $\hat{\bm{y}}_{jbk} = \text{SLM} \left( \Tilde{\bm{X}}_{jbk}^{tr}, \Tilde{\bm{X}}_{jbk}^{ts}, \bm{y}_b^{tr} \right)$.
                
                \item Calculate the accuracy, denoted as $\phi_{jbk}$, between $\bm{Y}_b^{ts}$ and $\hat{\bm{y}}_{jbk}$.
            \end{enumerate}
        \end{enumerate}
        
        \item Find the optimal parameter set, $\lambda_{k}^* = \lambda_{j_k^*}$, for $k$ using the average accuracy, $j_k^* = \argmax_{j} \frac{1}{B} \sum_{b=1}^{B} \phi_{jbk}$
        
        \item For each fold $b = 1, \ldots, B$,
        \begin{enumerate}[i.]\itemsep0pt 
            \item Cache the estimates  $\hat{\bm{v}}_{j_k^* b k}$, $\hat{\bm{w}}_{j_k^* b k}$, and $\hat{\bm{t}}_{j_k^* b k}$ using the optimal parameter set $\lambda_k^*$.
            
            \item Estimate $\hat{d}_{j_k^* b k}$ and deflate $\hat{\mathbfcal{Z}}_b$ as described by Eq. \eqref{eqn::deflation}, from Alg. \ref{alg::p-pca}.
            \vspace{2pt}
            \begin{equation*}
                \begin{split}
                \hat{d}_{j_k^* b k} &= \hat{\mathbfcal{Z}}_b \times_1 \hat{\bm{v}}_{j_k^* b k} \times_2 \hat{\bm{w}}_{j_k^* b k} \times_3 \hat{\bm{t}}_{j_k^* b k} \\
                \hat{\mathbfcal{Z}}_b &= \hat{\mathbfcal{Z}}_b - \hat{d}_{j_k^* b k} \cdot \hat{\bm{v}}_{j_k^* b k} \circ \hat{\bm{w}}_{j_k^* b k} \circ \hat{\bm{t}}_{j_k^* b k}
                \end{split}
            \end{equation*}
        \end{enumerate}
     \end{enumerate}
     \item Return the optimal parameters for each component, $\left\{ \left(\lambda_v^*, \lambda_w^*, \alpha_w^*, \alpha_t^* \right)_k \right\}_{k=1}^{K}$
     \item Run Alg. \ref{alg::p-pls} using the optimal parameters and full data set.
\end{enumerate}
\end{algorithm*}

\newpage
\bibliographystyle{apalike}
\bibliography{refs}

\end{document}